\documentclass[prd,aps,10pt,twocolumn,showpacs,reprint,nofootinbib]{revtex4}
 
\usepackage[dvips]{graphicx}
\usepackage{amssymb}
\usepackage{filecontents}
\usepackage{amsmath}
\usepackage{color}
\usepackage{float}
\usepackage{textcomp}
\usepackage{bm}

\RequirePackage[colorlinks,citecolor=blue,urlcolor=blue,linkcolor=blue]{hyperref}
\usepackage{yfonts}
\usepackage{amsmath,esint}
\usepackage{multirow}
\usepackage{enumerate}

\begin{document}
\title{Light bending and gravitational lensing in energy-momentum-squared gravity}
\author{Elham \surname{Nazari}$^{1}$}
\email{elham.nazari@mail.um.ac.ir}
\affiliation{$^1$Department of Physics, Faculty of Science, Ferdowsi University of Mashhad, P.O. Box 1436, Mashhad, Iran}

\begin{abstract}

In the present work, we derive the motion of light in the weak-field limit of energy-momentum-squared gravity (EMSG). To do so, we introduce the post-Newtonian (PN) expansion of this modified theory of gravity. It is shown that in addition to the Newtonian potential, a new EMSG potential affects the trajectory of photons. As a result, in this theory, photons do not behave as predicted by general relativity (GR).
To evaluate the EMSG theory by the solar system tests, we study light deflection and Shapiro time delay.
Regarding the results obtained in \cite{bertotti2003test,shapiro2004measurement}, we restrict the free parameter of the theory and show that it lies within the range $-4.0\times 10^{-27}\, \text{m}\,\text{s}^2\,\text{kg}^{-1} < f_0' < 8.7\times 10^{-26}\,\text{m}\,\text{s}^2\,\text{kg}^{-1}$.
This interval is in agreement with those derived in \cite{nazari2020Constraining,akarsu2018constraint}.   
This consistency manifests that this theory passes these solar system tests with flying colors. Interestingly, it turns out that the magnitude of the EMSG correction strongly depends on the density of the deflector.
So, we investigate the possible effects of EMSG on images of a light source microlensed by a compact dense object such as neutron stars. It is estimated that the EMSG correction to the position of lensed images could be as large as $(1-0.1)$ micro-arcseconds which may be detected by future high-resolution missions. 
Moreover, the total magnification and the shape of light curves are obtained in the EMSG theory. It is 
revealed that except for a small deviation, the overall behavior of the EMSG light curves is similar to that in GR.
We also show that as long as the light source and the dense lens are aligned, the EMSG correction is effective, and the combined light of the lensed images is different from the GR case. This issue makes it possible to observe signatures of this theory in the microlensing regime. 
\end{abstract}
\maketitle

\section{Introduction}

The energy-momentum-squared gravity (EMSG) is an alternative theory of gravity based on the action
principle. In addition to the usual Einstein-Hilbert term, the action incorporates the correction term $\boldsymbol{T}^2=T^{\alpha\beta}T_{\alpha\beta}$ built of the energy-momentum tensor, $T_{\alpha\beta}$, of the matter fields \cite{roshan2016energy,katirci2014f}.  
This new scalar term made up merely of the matter fields brings extra corrections to the right-hand side of the Einstein field equations. So, the matter fields are not conserved in this theory \cite{roshan2016energy}. 
In contrast to most higher-order theories of gravity referred to as $f(R)$ theories, where the gravitational Lagrangian is modified as a nonlinear function of the Ricci scalar curvature $R$, in EMSG, the higher order of the energy-momentum tensor of the matter fields is considered to modify the general relativity (GR).    
This theory has recently been taken into consideration and examined in several contexts \cite{board2017cosmological,akarsu2018cosmic,nari2018compact,akarsu2020screening,nazari2020generalized,barbar2020viability,kazemi2020jeans,sharif2021dynamics}. Moreover, applying some observational measurements, its free parameter has been constrained, e.g., see  \cite{akarsu2018constraint,nari2018compact,nazari2020Constraining}.

Meanwhile, it is necessary to check the validity of the theory by considering the local gravity as well as cosmological tests.  In this aspect, it is possible to set observational limits on the free parameter of the theory. 
It should be recalled that the deflection of light is one of the most powerful tools able to test modified theories of gravity at different scales. Light deflection and gravitational lensing have been studied in modified gravity theories. For instance, see \cite{tsujikawa2008effect,amendola2008measuring,moffat2009bending,capozziello2006gravitational}. 
Also, it has been shown that solar system data such as light bending can place strong constraints on parameters of $f(R)$ theories \cite{zakharov2006solar}.
In the case of EMSG, it is also possible that the modified gravitational Lagrangian leads to a change in the theory of light deflection and gravitational lensing. In fact, the effect of this modification may manifest itself in the motion of light. So, it is interesting to investigate the motion of light in the field of a source in EMSG and examine the EMSG corrections to gravitational lensing.  
 
In the present work, we derive the motion of light in the weak-field limit of EMSG. To do so, we introduce the post-Newtonian (PN) expansion of EMSG where slow-motion and weak gravitational field conditions are established. In \cite{nazari2020Constraining}, applying the modern approach to PN gravity \cite{will1996gravitational,pati2000post,pati2002post,poisson2014gravity}, the post-Minkowskian (PM) limit of this theory has been derived. Here, we employ the same technique to find the near-zone metric of the EMSG fluid up to the first PN (1\tiny PN \normalsize) order . The trajectory of photons in this spacetime is then obtained. As the first step to test this theory in the weak-field limit, we study light deflection and gravitational lensing by a spherically symmetric compact object in this work. 
We are interested in the possibility of constraining the free parameter of EMSG by studying light bending. As stellar lenses in the microlensing regime provide unambiguous measurements of light deflection by a compact object, we also attempt to find signatures of this theory in microlensing.

The paper is organized as follows.
The strategy of our calculations is clarified in Sec. \ref{The strategy of calculations}. As mentioned, this study is restricted to the PN limit of EMSG which is introduced in Sec. \ref{Post-Newtonian limit}. In this framework, each order $c^{-2}$ is considered as a PN correction. 
Appxes. \ref{app_1} and \ref{Newtonian hydrodynamics} are devoted to the comprehensive derivation of the PN expansion of this theory. In Sec. \ref{The motion of light}, we then find the motion of light in EMSG. In this section, the effect of the modification of gravity on light deflection, Shapiro time delay, and microlensing is examined. Finally, we conclude in Sec. \ref{Summary and Conclusion}. 

In this paper, Latin and Greek indices run over the values $\lbrace1,2,3\rbrace$ and $\lbrace0,1,2,3\rbrace$, respectively. Moreover, in our notation, $\eta_{\mu\nu}=\text{diag}(-1,1,1,1)$ is the Minkowski metric of flat spacetime and a spacetime event is labeled by $x^{\mu}=(c\,t,\boldsymbol{x})$.

\section{The strategy of calculations} \label{The strategy of calculations}

In this work, we focus on the motion of light in the vicinity of an EMSG source to examine signatures of the modification of gravity in the deflection of light and microlensing. The main goal is to study the motion of light in the weak-field limit of this theory. So, our first task is to build the PN limit of EMSG. 
To this aim, the modern approach to PN gravity is utilized. This method is based on the Landau-Lifshitz formulation of the gravitational theory \cite{poisson2014gravity}. In \cite{nazari2020Constraining}, this reformulation of the EMSG theory is derived comprehensively. In the following, we first mention the standard formulation of EMSG and then rewrite its Landau-Lifshitz one. 

The action of this theory is introduced as 
\begin{align}\label{EMSG-action}
S=\int \sqrt{-g}\Big(\frac{1}{2k}R+f_0'\, {\boldsymbol{T}}^2\Big)d^4x+S_{m},
\end{align}
in which $g$ is the determinant of the spacetime metric $g_{\mu\nu}$, $k=8\pi G/c^4$, $S_m$  is the matter action, and $R$ is the spacetime curvature. Here, $f_0'$ is the free parameter of the theory representing the coupling between matter and spacetime
 \footnote{In this work, we adopt the same notation applied for the free parameter of the EMSG theory in \cite{nazari2020Constraining}.}.
The EMSG field equations are derived in \cite{roshan2016energy}. In our notation, the field equations are given by
\begin{align}\label{G}
G_{\mu\nu}=
k\Big( T_{\mu\nu}+f_0'\big( g_{\mu\nu}{\boldsymbol{T}}^2-4T^\sigma_\mu T_{\nu\sigma}-4{\boldsymbol{\Psi}}_{\mu\nu}\big)\Big),
\end{align}
where $G_{\mu\nu}$ is the Einstein tensor and   
\begin{align}
{\boldsymbol{\Psi}}_{\mu\nu}=-L_m\big(T_{\mu\nu}-&\frac{1}{2}Tg_{\mu\nu}\big)\\\nonumber
&-\frac{1}{2}TT_{\mu\nu}-2T^{\alpha\beta}\frac{\partial^2 L_m}{\partial g^{\alpha\beta}\partial g^{\mu\nu}}.
\end{align}
Here, $L_m$ stands for the matter Lagrangian density and $T$ is the trace of the energy-momentum tensor.
It should be mentioned that the matter Lagrangian density is independent of metric derivatives and it is only a function of the metric. 
Obviously due to the extra terms on the right-hand side of Eq. \eqref{G}, the usual energy-momentum tensor is not conserved in this gravitational theory, i.e., $\nabla_{\mu}T^{\mu\nu}\neq 0$. Given the Bianchi identities, one can deduce that instead of $T_{\mu\nu}$, the effective energy-momentum tensor $T_{\mu\nu}^{\text{eff}}$ defined as $T_{\mu\nu}+f_0'\big( g_{\mu\nu}{\boldsymbol{T}}^2-4T^\sigma_\mu T_{\nu\sigma}-4{\boldsymbol{\Psi}}_{\mu\nu}\big)$ would be conserved here. So, we have
\begin{align}\label{nabla_T_eff}
\nabla_{\mu}T^{\mu\nu}_{\text{eff}}=0.
\end{align}
It should also be mentioned that for different Lagrangian densities
that describe a perfect fluid, the EMSG field equations \eqref{G} would be inequivalent. Therefore, for various Lagrangian densities, this theory makes different predictions. This issue is investigated in other modified theories of gravity, cf. \cite{faraoni2009lagrangian} and references therein. Throughout this work, we utilize the standard Lagrangian density $L_m=p$ for a perfect fluid.
During our calculation, we also assume that the mass-current vector, $\rho\, u^{\mu}$, is divergence-free in this theory, i.e., $\nabla_{\mu}\big(\rho u^{\mu}\big)=0$ where $\rho$ is the rest-mass density and $u^{\mu}=\gamma(c,\boldsymbol{v})$ is the four-velocity field. Here, $\gamma=u^0/c$ and $\boldsymbol{v}=d\boldsymbol{x}/dt$ is the three-velocity field.

In \cite{nazari2020Constraining}, we have shown that the Landau-Lifshitz reformulation of the EMSG field equations is simplified as
\begin{align}\label{wave_eq}
\square h^{\mu\nu}=-2k\tau^{\mu\nu}_{\text{eff}},
\end{align}
where $\square=-\frac{1}{c^2}\frac{\partial^2}{\partial t^2}+\nabla^2$ is the wave operator in the flat spacetime and  $h^{\mu\nu}$ is the gravitational potential defined as $\eta^{\mu\nu}-\sqrt{-g}g^{\mu\nu}$. Here,  $\tau^{\mu\nu}_{\text{eff}}=(-g)\Big(T^{\mu\nu}_{\text{eff}}+t_{\text{LL}}^{\mu\nu}+t_{\text{H}}^{\mu\nu}\Big)$ is the effective energy-momentum pseudotensor written in terms of $T^{\mu\nu}_{\text{eff}}$, the Landau-Lifshitz, $(-g)t_{\text{LL}}^{\mu\nu}$, and harmonic, $(-g)t_{\text{H}}^{\mu\nu}$, pseudotensors. The definition of these pseudotensors is given in Appx. \ref{app_1}. See Eqs. \eqref{tLL} and \eqref{tH}.
In this formulation, it is assumed that the gravitational potential $h^{\mu\nu}$ satisfies the harmonic gauge conditions $\partial_{\mu}h^{\mu\nu}=0$.

In the following, to derive the PN approximation to the EMSG field equations, we approximately solve the highly non-linear wave equation \eqref{wave_eq}. It should be emphasized that in \cite{nazari2020Constraining}, to study the gravitational radiation effects in this theory, Eq. \eqref{wave_eq} is solved in the wave zone, while in order to investigate the motion of light in the vicinity of an EMSG lens, we restrict ourselves to the near-zone solutions \footnote{In the next section, the near and wave zones are introduced in the framework of PN gravity.} in the current work. It should also be noted that this wave equation is similar to the GR case, cf. \cite{poisson2014gravity}. As seen, an explicit difference comes from the additional EMSG terms within $T^{\mu\nu}_{\text{eff}}$ on the right-hand side of this relation. However, we will see that some implicit differences arise from the rest source terms, i.e., $t_{\text{LL}}^{\mu\nu}$ and $t_{\text{H}}^{\mu\nu}$.  
This similarity with the GR case in fact allows us to employ the same techniques introduced in \cite{poisson2014gravity} to solve this wave equation.
We recall that contrary to GR, $T^{\mu\nu}$ is not conserved here. So, these techniques should be applied with careful treatments.
Until now, we do not apply any approximation and Eq. \eqref{wave_eq} is only the reformulation of the EMSG field equations after imposing the harmonic gauge conditions.

The final point here is to answer this question: to what PN order should we carry out our calculations and solve this wave equation? As mentioned earlier, our goal is to test EMSG theory in the weak deflection limit where the light ray's distance of closest approach, $b$, lies far beyond the gravitational radius, $2G M/c^2$, of the lens with mass $M$.  
Therefore, we need to know the spacetime geometry far from the gravitational radius of the lens, where 1\tiny PN \normalsize corrections would be sufficient and higher PN orders can be freely ignored. We recall that in the PN framework, each order $c^{-2}$ is considered as a PN correction. 
Furthermore, in \cite{nazari2020Constraining}, using a crude estimation, it has been shown that EMSG corrections should be considerably small.
It has been revealed that the EMSG parameter $f_0'$ is at most of the order of $c^{-2}$. Here, we apply this estimation for the magnitude of the EMSG parameter and treat each $f_0'$ order as a PN correction. 
During our derivations, we will see that this is a reasonable assumption.
Given the above points, the spacetime metric of an EMSG fluid is comprehensively derived up to the 1\tiny PN \normalsize corrections in Appx. \ref{app_1}. The trajectory of photons in this spacetime is obtained in the following section.

\section{Post-Newtonian limit of EMSG}\label{Post-Newtonian limit}

In this section, we find the near-zone metric of the EMSG fluid up to the 1\tiny PN \normalsize order.
To do so, we first need to obtain the relation between the components of the metric and those of the gravitational potential $h_{\mu\nu}$. In the framework of the modern approach, this relation is derived in \cite{nazari2020Constraining}. Next, in order to find $h_{\mu\nu}$ and consequently construct the components of $g_{\mu\nu}$, we should solve the wave equation \eqref{wave_eq}. 
Detailed discussions and calculations are provided in Appx. \ref{app_1}.

In the context of the PN approximation, the near and wave zones are respectively the region inside and outside of a sphere with the radius $\mathcal{R}\approx\lambda_{\text{c}}$ in which $\lambda_{\text{c}}$ is the characteristic wavelength of the gravitational signals generated by the system.   
Moreover, the slow-motion condition, ${v_\text{c}}^2/c^2\ll 1$, and weak-field limit, $U/c^2\ll 1$, are the two essential conditions that are fulfilled in the PN limit. Here, $v_\text{c}$ is a characteristic velocity within the fluid and $U$ is the Newtonian potential. In the modern approach to the PN limit of GR, regarding the position of the field point and applying these conditions as well as the iterative procedure, the wave equations/Einstein field equations are approximately solved to the required degree of accuracy \cite{poisson2014gravity}. Here, in a similar manner to the GR case, using the iterative method \footnote{We refer readers who are unfamiliar with the iterative procedure in the modern approach to chapters 6 and 7 of \cite{poisson2014gravity}.}, we approximately solve the EMSG field equations \eqref{wave_eq} where the field point is located in the near-zone region of spacetime.

Before introducing the near-zone solutions, let us review the PN order of the metric components we need to study a system at least up to the 1\tiny PN \normalsize order. As we know, to obtain the 1\tiny PN \normalsize correction to the equation of motion of a test particle, we should study the Lagrangian up to order $c^{-2}$. Therefore, the time-time, space-time, and space-space components of the metric should be evaluated up to $O(c^{-4})$, $O(c^{-3})$, and $O(c^{-2})$, respectively. On the other hand, in order to obtain the PN corrections to the propagation of light rays, only $O(c^{-2})$ must be known for both time-time and space-space components of the metric. However, for completeness, we derive all PN terms required to study a system up to the 1\tiny PN \normalsize order.
In \cite{nazari2020Constraining}, the general expansion of the metric components in terms of the gravitational potential $h_{\mu\nu}$ for a perfect fluid has been found, see Eqs. (29a)-(29c) and (30) in this reference. For the sake of convenience, we rewrite these expansions here.
\begin{subequations}
\begin{align}
\nonumber
&
 g_{00}=-1+\frac{1}{2}h^{00}-\frac{3}{8}\big(h^{00}\big)^2+\frac{1}{2}h^{kk}\big(1-\frac{1}{2}h^{00}\big)\\\label{g00}
&-\frac{1}{8}\big(h^{kk}\big)^2+O(c^{-6}),\\
\label{g0j}
& g_{0j}=-h^{0j}+O(c^{-5}),\\
\label{gjk}
& g_{ij}=\delta_{ij}\big(1+\frac{1}{2}h^{00}\big)+h^{ij}-\frac{1}{2}\delta_{ij}h^{kk}+O(c^{-4}),\\\label{g}
&(-g)=1+h^{00}-h^{kk}+O(c^{-4}).
\end{align}
\end{subequations}
In the above relations, $\delta_{ij}$ is the Kronecker delta and $h^{kk}$ is the trace of $h^{jk}$.
Considering the PN order of the EMSG gravitational potential components evaluated in Appx. \ref{app_1}, these expansions would in fact provide the required PN order for the metric components stated previously. 
In other words, each term in the above relations will contribute to the 1\tiny PN \normalsize correction of the equation of motion of a test particle in the EMSG spacetime.

We now turn to obtain the solution of the wave equation \eqref{wave_eq}.  As mentioned, using the iteration method, we solve this highly non-linear equation in the EMSG theory. 
Here, we choose a perfect fluid whose energy-momentum tensor is described by $T^{\mu\nu}=(\rho+\epsilon/c^2+p/c^2)u^{\mu}u^{\nu}+pg^{\mu\nu}$. In the framework of the PN limit, 
we assume that the proper energy density $\epsilon$ and the pressure $p$ satisfy the two conditions $\epsilon/\rho c^2\ll 1$ and $p/\rho c^2\ll 1$, respectively. Similar to the GR case, we should also carry out our calculations up to the second iterated step to achieve the desired PN corrections for building the 1\tiny PN \normalsize order of the near-zone metric. As this derivation is long and also tedious, we remove this part from the main text and add the complete calculations to Appx. \ref{app_1}.

\subsection{Near-zone metric} \label{near-zone metric}

Here, we apply the final relations for the time-time, space-time, and space-space components of $h_{\mu\nu}$ obtained in Appx. \ref{app_1}.
Substituting Eqs. \eqref{h002}, \eqref{hkk2}, \eqref{h0j2}, and  \eqref{eq3} into Eqs.  \eqref{g00}-\eqref{g}, we arrive at 
\begin{subequations}
\begin{align}
\label{G00}
& g_{00}=-1+\frac{2}{c^2}U+\frac{2}{c^4}\Big(\psi+V-U^2+\frac{1}{2}\partial_{tt} X\Big)\\\nonumber
&+4f'_0\Big[2U_{\text{\tiny EMS}}+\frac{1}{c^2}\Big(\partial_{tt}X_{\text{\tiny EMS}}+4P_{\text{\tiny EMS}}-4UU_{\text{\tiny EMS}}\\\nonumber
&-8\,\mathcal{U}^{(1)}_{\text{\tiny EMS}}-2\,\mathcal{U}^{(2)}_{\text{\tiny EMS}}+4\Pi_{\text{\tiny EMS}}\Big)-11f'_0U^2_{\text{\tiny EMS}}\Big]+O(c^{-6}),\\
& g_{0j}=-\frac{4}{c^3}U_j-\frac{8f'_0}{c}U_{j\text{\tiny EMS}}+O(c^{-5}),\\
\label{Gjk}
& g_{jk}=\delta_{jk}\Big(1+\frac{2}{c^2}U\Big)+O(c^{-4}),
\end{align}
\end{subequations}
and 
\begin{align}
\big(-g\big)=1-8f'_0U_{\text{\tiny EMS}}+\frac{4}{c^2}U+O(c^{-4}),
\end{align}
for the components of the metric and its determinant, respectively.
This near-zone metric describes the spacetime of the PN perfect fluid in the EMSG theory. As it is seen, in addition to the well-known PN potentials, i.e., $\psi$, $V$, $X$, $\boldsymbol{U}$, one needs extra gravitational potentials indicated with the index EMS, to correctly study the behavior of a system in the weak-field limit of the EMSG theory. The definition of these potentials is given in Appx. \ref{app_1}, see  Eqs. \eqref{U_EMSG}, \eqref{UjEMSG}, \eqref{XEMS}, and \eqref{Pi}-\eqref{U22}.
Most of these new PN potentials are constructed from the matter part of the system. The source term of an EMSG potential, which we call the EMSG superpotential $X_{\text{\tiny EMS}}$, itself is a gravitational potential extending overall space. Given the Poisson integrals \eqref{U_EMSG} and \eqref{XEMS}, one can show that this superpotential satisfies the Poisson equation $\nabla^2X_{\text{\tiny EMS}}=2U_{\text{\tiny EMS}}$. This in fact is equivalent to $X$ in the well-known PN expansion of the metric in GR. Considering the coefficients $c^{-2}$ and $f_0'$, one can deduce that these components of the metric are truly constructed up to the desired PN order mentioned earlier.

We have now enough information to study the dynamics of a system/motion of light with the 1\tiny PN \normalsize corrections in the EMSG theory. The remainder of this section is devoted to the computation of the photon's trajectory in the EMSG curved spacetime.

\subsection{Geodesic equations}\label{Motion of a photon in EMSG}

We consider the geometric-optics approximation where massless particles/photons can describe the behavior of light rays. We first choose a pressure-less perfect fluid described by
\begin{align}
T^{\mu\nu}\vert_{\text{dust}}=\rho\, u^{\mu}u^{\nu}.
\end{align}
One can show that for this fluid, the EMSG part of $T^{\mu\nu}_{\text{eff}}$ is given by
\begin{align}
T^{\mu\nu}_{\text{\tiny EMS}}\vert_{\text{dust}}=f'_0\Big(c^4\rho^2g^{\mu\nu}+2c^2\rho^2u^{\mu}u^{\nu}\Big).
\end{align}
To simplify this relation, we use $u_{\mu}u^{\mu}=-c^2$.
Inserting the above relations into Eq. \eqref{nabla_T_eff} and using the conservation of rest-mass $\nabla_{\mu}\big(\rho u^{\mu}\big)=0$, we arrive at
\begin{align}\label{traject}
u^{\mu}\nabla_{\mu}u^{\nu}=-\frac{2c^2f'_0\big(c^2g^{\mu\nu}+u^{\mu}u^{\nu}\big)\partial_{\mu}\rho}{1+2c^2f'_0\rho}.
\end{align}
Eq. \eqref{traject} reveals that in the EMSG theory, the worldline of a dust particle is not necessarily described by the geodesic equation. The non-zero term on the right-hand side of this relation can be interpreted as an additional force acting on particles that prevents them from moving on the geodesic of spacetime. A similar issue is also pointed out in the Palatini formalism of the generalized EMSG in \cite{nazari2020generalized}.

In a similar fashion, for the null dust fluid, we find
\begin{align}\label{geo}
u^{\mu}\nabla_{\mu}u^{\nu}=0.
\end{align}
To simplify the above relation, the null condition $u_{\mu}u^{\mu}=0$ is utilized. Since the right-hand side of Eq. \eqref{geo} is zero, unlike the previous case, the worldline of a massless particle/photon is a geodesic. 
As usual, this relation can be simplified as follows:
\begin{align}
\label{geo1}
\frac{d^2 x^{\nu}}{d\lambda^2}+\Gamma^{\nu}_{\alpha\beta}\frac{dx^{\alpha}}{d\lambda}\frac{dx^{\beta}}{d\lambda}=0,
\end{align}
where the quantities $\Gamma^{\nu}_{\alpha\beta}$ are the Christoffel symbols.
Here, we consider that $x^{\nu}(\lambda)$ is the worldline of photons parametrized by an arbitrary affine parameter $\lambda$. One can show that by changing $\lambda$ to the time coordinate $t$, this relation reduces to
\begin{align}\label{geo2}
\frac{dv^{\nu}}{dt}+\Big(\Gamma^{\nu}_{\alpha\beta}-\frac{v^{\nu}}{c}\Gamma^{0}_{\alpha\beta}\Big)v^{\alpha}v^{\beta}=0,
\end{align}
where $v^{\mu}=dx^{\mu}/dt$.

In order to find the photon's trajectory in a PN EMSG fluid, we now apply the near-zone metric built up to the 1\tiny PN \normalsize order in the previous section.
Obtaining the Christoffel symbols and after algebraic simplification, we arrive at
\begin{widetext}
\begin{align}
\nonumber
&\frac{dv^j}{dt}-\partial_jU-4c^2f'_0\partial_jU_{\text{\tiny EMS}}-\frac{1}{c^2}\bigg\lbrace \big(v^2-4U\big)\partial_jU-\big(4v^k\partial_kU+3\partial_tU\big)v^j-4v^k\big(\partial_jU_k-\partial_kU_j\big)+4\partial_tU_j+\partial_j\Psi\\\nonumber
&+8c^2f'_0\Big[v^k\big(\partial_kU_{j}^{\text{\tiny EMS}}-\partial_jU_{k}^{\text{\tiny EMS}}\big)-\big(v^k\partial_kU_{\text{\tiny EMS}}+\frac{1}{2}\partial_tU_{\text{\tiny EMS}}\big)v^j+\frac{1}{4}\partial_j\partial_{tt}X_{\text{\tiny EMS}}+\partial_jP_{\text{\tiny EMS}}+\partial_j\Pi_{\text{\tiny EMS}}-U_{\text{\tiny EMS}}\partial_jU-\big(2U\\\label{geo3}
&+\frac{11}{2}c^2f'_0U_{\text{\tiny EMS}}\big)\partial_jU_{\text{\tiny EMS}}-2\partial_j\mathcal{U}^{(1)}_{\text{\tiny EMS}}-\frac{1}{2}\partial_j\mathcal{U}^{(2)}_{\text{\tiny EMS}}+\partial_tU_j^{\text{\tiny EMS}}\Big]\bigg\rbrace+O(c^{-4})=0,
\end{align}
\end{widetext}
for the geodesic equation \eqref{geo2}. Here, $\Psi=\psi+V+1/2\partial_{tt}X$. We then use the lightlike condition $g_{\alpha\beta}v^{\alpha}v^{\beta}=0$ to evaluate the PN order of each term in this relation. 
Using Eqs. \eqref{G00}-\eqref{Gjk} and expanding this condition, one can deduce that
\begin{align}\label{condi}
v^j=c\Big(1-\frac{2U}{c^2}-4f'_0U_{\text{\tiny EMS}}\Big)n^j+O(c^{-3}),
\end{align}
in which $\boldsymbol{n}$ is a unit vector that indicates the direction of light emission. This relation shows that the coordinate velocity of a photon is of the order of $c$.
Keeping this in mind, we neglect the small terms from Eq. \eqref{geo3}. We then have
\begin{align}\label{eq4}
\nonumber
\frac{dv^j}{dt}=&\Big(1+\frac{v^2}{c^2}\Big)\partial_jU+4c^2f'_0\partial_jU_{\text{\tiny EMS}}-\frac{4}{c^2}\Big[v^jv^k\big(\partial_kU\\
&+2c^2f'_0\partial_kU_{\text{\tiny EMS}}\big)\Big]+O(c^{-3}).
\end{align}
Substituting Eq. \eqref{condi} into Eq. \eqref{eq4} gives
\begin{align}
\label{geonull}
\frac{dn^j}{dt}=\frac{2}{c}\Big(\delta^{jk}-n^jn^k\Big)\Big(\partial_kU+2c^2f'_0\partial_kU_{\text{\tiny EMS}}\Big)+O(c^{-3}).
\end{align}
To achieve this expression, we use that $dU/dt\simeq c\, n^k\partial_kU$ and $dU_{\text{\tiny EMS}}/dt\simeq c\,n^k\partial_kU_{\text{\tiny EMS}}$. It is obvious that dropping EMSG correction, this relation reduces to the GR case. 
Eq. \eqref{geonull} is the null geodesic with which we study the motion of light in the weak-field limit of the EMSG theory.

\section{The motion of light}\label{The motion of light}

We launch this section by solving Eq. \eqref{geonull}.
Given the PN order of this differential equation, its solution up to the leading order can be written as $n^j=k^j+O(c^{-2})$ where $k^j$ is the $j$th component of a constant vector $\boldsymbol{k}$. Then, substituting this relation back within Eq. \eqref{condi}, gives $v^j=c\,k^j+O(c^{-1})$ for the  photon's coordinate velocity. Next, solving the differential equation $dx^j/dt=c\,k^j+O(c^{-1})$, we obtain 
$\boldsymbol{x}(t)=\boldsymbol{x}_{\text{e}}+c\,\boldsymbol{k}\big(t-t_{\text{e}}\big)+O(c^{-2})$. Here, $\boldsymbol{x}_{\text{e}}$ is the position of the photon at the emission time $t_{\text{e}}$. Up to this order, the photon moves in a straight path. We continue our calculation until the gravitational effects, especially those related to the EMSG theory, appear in the photon's trajectory.

We consider the next order of light direction as $n^j=k^j+\alpha^j+O(c^{-4})$ in which $\alpha^j$ is the 1\tiny PN \normalsize correction to $n^j$. Regarding Eq. \eqref{geonull}, this correction should satisfy the following relation
\begin{align}
\label{dn_j}
\frac{d\alpha^j}{dt}=\frac{2}{c}\Big(\delta^{jk}-k^jk^k\Big)\Big(\partial_kU+2c^2f'_0\partial_kU_{\text{\tiny EMS}}\Big).
\end{align}
This relation illustrates that $\boldsymbol{\alpha}.\boldsymbol{k}=0$. Since the EMSG correction appears in this parameter, henceforth we call it $\alpha_\text{{\tiny EMS}}$. 
In this order, we then have
\begin{align}
v^j=c\Big(1-\frac{2U}{c^2}-4f'_0U_{\text{\tiny EMS}}\Big)k^j+c\, \alpha_{\text{\tiny EMS}}^j+O(c^{-3}),
\end{align}
for Eq. \eqref{condi}.
Finally, for the photon's trajectory, we arrive at 
\begin{align}
\boldsymbol{x}=\boldsymbol{x}_{\text{e}}+c\,\boldsymbol{k}\big(t-t_{\text{e}}\big)+\boldsymbol{k}\,x_{\shortparallel}(t)+\boldsymbol{x}_{\perp}(t)+O(c^{-4}),
\end{align}
where 
\begin{subequations}
\begin{align}
\label{eqshap}
& x_{\shortparallel}(t)=-2\int_{t_{\text{e}}}^{t}\Big(\frac{U}{c}+2c f_0'U_{\text{\tiny EMS}}\Big)dt',\\
& \boldsymbol{x}_{\perp}(t)=c\int_{t_{\text{e}}}^{t}\boldsymbol{\alpha}_{\text{\tiny EMS}}\,dt',
\end{align}
\end{subequations}
indicate the longitudinal and transverse corrections to the trajectory of photons, respectively. As seen, both of these relativistic corrections include the EMSG terms.
In fact, in this theory, in addition to the Newtonian potential, the EMSG potential $U_{\text{\tiny EMS}}$ affects both longitudinal and transverse terms in the trajectory of photons. As a result, the photon does not behave as predicted by GR.
In the following subsections, to understand if there is a detectable signature of the EMSG theory in the motion of light, we examine each of these corrections.

\subsection{Light deflection} \label{LD}

Here, we study light deflection by a spherically symmetric compact object.
In order to derive the deflection angle, we should find the gravitational potentials $U$ and $U_{\text{\tiny EMS}}$ induced by this object.  
To do so, we place the origin of the coordinate system in the center of the body.
Our goal is to find these potentials outside of this body where the 1\tiny PN \normalsize approximation works well. So, the position of the field point would be larger than the dimension of the compact object and as a result, we can simplify Eqs. \eqref{U} and \eqref{U_EMSG} as \footnote{Here, as our aim is to find the leading order of light deflection in the EMSG theory, we have exhibited the mass density with $\rho$ instead of $\rho^*$ introduced in Appx. \ref{app_1}.} 
\begin{subequations}
\begin{align}
\label{U_body}
& U=\frac{G\, M}{r}+O(r^{-3}),\\
\label{UE_body}
& U_{\text{\tiny EMS}}=\frac{G \,\mathfrak{M}}{r}+O(r^{-3}),
\end{align}
\end{subequations}
where $r=|\boldsymbol{x}|$ and
\begin{subequations}
\begin{align}
& M =\int_{V_{\text{b}}}\rho\, d^3x,\\
& \mathfrak{M}=\int_{V_{\text{b}}}\rho^2d^3x,
\end{align}
\end{subequations}
are the material mass and the EMSG parameter of the body, respectively. Here, $V_{\text{b}}$ shows the volume occupied by this body. Substituting the gravitational potentials \eqref{U_body} and \eqref{UE_body} back within Eq. \eqref{dn_j}, one can arrive at 
\begin{align}
\frac{d\alpha^j_{\text{\tiny EMS}}}{dt}=-\frac{2\,G}{c}\big(M+2c^2f_0' \mathfrak{M}\big)\frac{b^j}{r^3},
\end{align}
in which $b^j=x_{\text{e}}^j-k^j(\boldsymbol{k}.\boldsymbol{x}_{\text{e}})$ and $\boldsymbol{x}(t)=\boldsymbol{x}_{\text{e}}+c\boldsymbol{k}\big(t-t_{\text{e}}\big)+O(c^{-2})$. To simplify this relation, we consider that the density of the body is constant, i.e., $\boldsymbol{\nabla} \rho=0$. So, given the point mentioned in Appx. C of \cite{nazari2020Constraining}, we can easily set $d\mathfrak{M}/dt=0$. Keeping this fact in mind and knowing that $d\big(\boldsymbol{b}/b^2\big)/dt=0$ as well as  $c\, b^2/r^3=d(\boldsymbol{k}.\boldsymbol{x}/r)/dt$, we obtain 
\begin{align}
\alpha^j_{\text{\tiny EMS}}=-\frac{2G}{c^2}M\big(1+2c^2f_0'\rho\big)\frac{b^j}{b^2}\Big(\cos\Phi(t) +1\Big).
\end{align}
Here, $b=|\boldsymbol{b}|$ and $\cos \Phi(t)=\boldsymbol{k}.\boldsymbol{x}/r$. It should be mentioned, in the above relation, we assume that $\boldsymbol{\alpha}_{\text{\tiny EMS}}(t_{\text{e}})=0$ and $b\ll r_{\text{e}}$. Similar to the GR case, we also consider a simple case where $\boldsymbol{k}.\boldsymbol{x}/r=1$ or $t \rightarrow \infty$. Under this circumstance, we find
\begin{align}\label{lighdef}
\boldsymbol{\alpha}_{\text{\tiny EMS}}=-\frac{4G}{c^2}M\Big(1+2c^2f'_0\rho\Big)\frac{\boldsymbol{b}}{b^2},
\end{align}
for the deflection of light from a sphere with a constant density $\rho$ and mass $M$ in the EMSG theory. Obviously, by dropping the EMSG correction, this formula reduces to light deflection in the weak field limit of Einsteinian relativity and its classical results are recovered. 
As seen, in addition to the mass of the system, $\alpha_{\text{\tiny EMS}}$ depends on its density. Also, compared to GR, the deflection of light in EMSG takes different values depending on the value of $f_0'$.
For the positive/negative value of $f_0'$, the EMSG version of the deflection of light can be more/less than the GR case.
Thus, EMSG may leave an observational signature on the deflection of light from a dense compact system. 

We utilize the measurement of light deflection introduced in the literature to find a reasonable bound on the free parameter of the EMSG theory. To do so, we consider that the EMSG version of the light deflection can fully describe the observed deflection by the Sun. Given this point and using our results and those in \cite{shapiro2004measurement}, we set $\alpha_{\text{\tiny EMS}}=\alpha_{\text{obs}}$. We then have $\gamma=1+4c^2f_0'\rho_{\odot}$ where $\rho_{\odot}$ is the density of the Sun and $\gamma$ is the parametrized post-Newtonian (PPN) parameter exhibiting the role of space curvature in the gravitational deflection. Regarding the numerical value of this PPN parameter obtained in \cite{shapiro2004measurement}, i.e., $\gamma=0.9998\pm 0.0004$, one can find the interval $-1.2\times 10^{-24}\, \text{m}\,\text{s}^2\,\text{kg}^{-1} < f_0' < 4.0\times 10^{-25}\,\text{m}\,\text{s}^2\,\text{kg}^{-1}$. Here, $\rho_{\odot}=1.4\, \text{g}\,\text{cm}^{-3}$. Therefore, the free parameter of this theory should lie within this domain to justify this solar system test. Although this range is not more limited than the previous studies \footnote{By studying neutron stars and binary pulsars, \cite{akarsu2018constraint} and \cite{nazari2020Constraining} show that the free parameter of the theory lies within the range $-10^{-37}\, \text{m}\,\text{s}^2\,\text{kg}^{-1} < f_0' < + 10^{-36}\,\text{m}\,\text{s}^2\,\text{kg}^{-1}$.}, there is no inconsistency with them and it well covers the intervals obtained in \cite{nazari2020Constraining,akarsu2018constraint}. This means that the EMSG theory passes this solar system test with flying colors.

\subsection{Shapiro time delay}\label{STD}

The Cassini tracking measurement of the Shapiro time delay can put another empirical constraint on the free parameter of the modified theories of gravity. In this subsection, benefiting this measurement, we attempt to obtain an experimental bound on $f_0'$. 

To do so, we first derive the EMSG correction to the Shapiro time delay. As seen, the longitudinal correction to the trajectory of photons is described by Eq. \eqref{eqshap} in the EMSG theory.
In a similar fashion to the method utilized in the previous subsection, we consider a spherically symmetric compact object as our gravitational system and then insert Eqs. \eqref{U_body} and \eqref{UE_body} into the definition \eqref{eqshap}. For the sake of simplification, we also assume that the density of the compact system is uniform. Regarding this point, we have 
\begin{align}
 x_{\shortparallel}(t)=-\frac{2 G}{c}M\Big(1+2c^2 f_0' \rho\Big)\int_{t_{\text{e}}}^{t_{\text{obs}}}\frac{1}{r}\,dt'.
\end{align}
Given $c/r=d\big(\ln(r+\boldsymbol{k}.\boldsymbol{x}) \big)/dt$, we then arrive at
\begin{align}\label{EQ1}
 x_{\shortparallel}(t)=-\frac{2G}{c^2}M\Big(1+2c^2f_0'\rho\Big)\ln\bigg[\frac{4 r_{\text{obs}} r_{\text{e}}}{b^2}\bigg],
\end{align}
for the EMSG longitudinal correction to the trajectory of photons. Here, ${r}_{\text{obs}}={r}(t=t_{\text{obs}})$. We also assume the case where $\boldsymbol{k}.\boldsymbol{x}_{\text{obs}}\simeq r_{\text{obs}}$ and $\boldsymbol{k}.\boldsymbol{x}_{\text{e}}\simeq-r_{\text{e}}$. According to the relation \eqref{EQ1}, one can deduce that the Shapiro time delay is given by
\begin{align}\label{shapiro}
\Delta t_{ \text{\tiny Shapiro}}^{\text{\tiny EMS}}=\frac{4G}{c^3}M\Big(1+2c^2f_0'\rho\Big)\ln\bigg[\frac{4 r_{\text{obs}} r_{\text{e}}}{b^2}\bigg],
\end{align}
in the EMSG theory. Comparing Eqs \eqref{lighdef} and \eqref{shapiro} reveals that the EMSG contribution to the light deflection and Shapiro time delay is similar and in both cases, the same expression $\big(1+2c^2f_0'\rho\big)$ is added. 
This relation also indicates that for the system with high density where gravity experiments are carried out, the EMSG effects can be significant or even detectable.

Here, to find an experimental bound on $f_0'$, we apply the results of \cite{bertotti2003test}. It is assumed that this theory can truly justify the measurement of the Shapiro time delay in the solar-system situation. So, we set $\Delta t_{ \text{\tiny Shapiro}}^{\text{\tiny EMS}}=\Delta t_{ \text{\tiny Shapiro}}^{\text{\tiny obs}}$. Using $\gamma=1+(2.1\pm 2.3)\times 10^{-5}$ \cite{bertotti2003test}, we find that $-4.0\times 10^{-27}\, \text{m}\,\text{s}^2\,\text{kg}^{-1} < f_0' < 8.7\times 10^{-26}\,\text{m}\,\text{s}^2\,\text{kg}^{-1}$. This bound is also in agreement with the previous studies \cite{akarsu2018constraint,nazari2020Constraining}. Therefore, this theory is not ruled out by this solar system test. It is worthwhile to mention that the boundary obtained here is tighter than that inferred from light deflection. This is because the standard error of the PPN parameter $\gamma$ in \cite{bertotti2003test} is smaller than that in \cite{shapiro2004measurement}. Therefore, as expected, by improving the measurement of the Shapiro time delay and light deflection in the solar system, the free parameter of this theory can be even more restricted.

\subsection{Gravitational lensing}

As seen, the density of the compact system directly appears in the definition of light deflection and  Shapiro time delay, and the magnitude of the EMSG correction strongly depends on it. This is also the case in \cite{nazari2020Constraining} where the EMSG correction to the orbital period variation of binary pulsars is studied.  As a result, for the dense system, the EMSG effect can be significant even if the order of magnitude of the parameter $f_0'$ is very small. In fact, the denser the system, the more important the EMSG corrections are. 
On the other hand, the gravitational system considered here is the Sun, which has a very low density compared to neutron stars studied in \cite{nazari2020Constraining,akarsu2018constraint} to constrain $f_0'$.  This is the main reason why the interval obtained in the earlier subsections \ref{LD} and \ref{STD} is not more limited than what is introduced in these papers.

In this subsection, to examine possible detectable signatures of EMSG in light curves, we focus our attention on the gravitational lensing from a dense compact object like neutron stars. To achieve this goal, we first derive the lens equation in the EMSG theory. Regarding the spherical symmetry of the lens considered here and applying the small-angle approximation, we have 
\begin{align}
\theta-\frac{d_{\text{ls}}}{d_{\text{so}}}\alpha_{\text{\tiny EMS}}=\theta_{\text{s}},
\end{align}
which is the scalar form of the lens equation written in the deflector's plane. The EMSG effects are encoded in $\alpha_{\text{\tiny EMS}}$ in this equation. Here, $\theta$ and $\theta_{\text{s}}$ stand for the angle between the image(s)-optical axis and the source-optical axis, respectively. Note that the line connecting the observer to the lens is defined as the optical axis, and the lens and source planes are perpendicular to the initial direction of the photon path. Also, $d_{\text{ls}}$ and $d_{\text{so}}$ show the angular
diameter distances between lens-source and source-observer, respectively. It is worth pointing out that in the following calculations, the angular diameter distances are postulated to be well-defined in GR. These distances are not clearly defined in this theory. Nevertheless, for a full analysis, it is necessary to obtain these distances and their relationship to redshift in EMSG. We leave this for the future.

%================================================================================
\begin{figure}
\begin{center}
\centering
\includegraphics[scale=0.63]{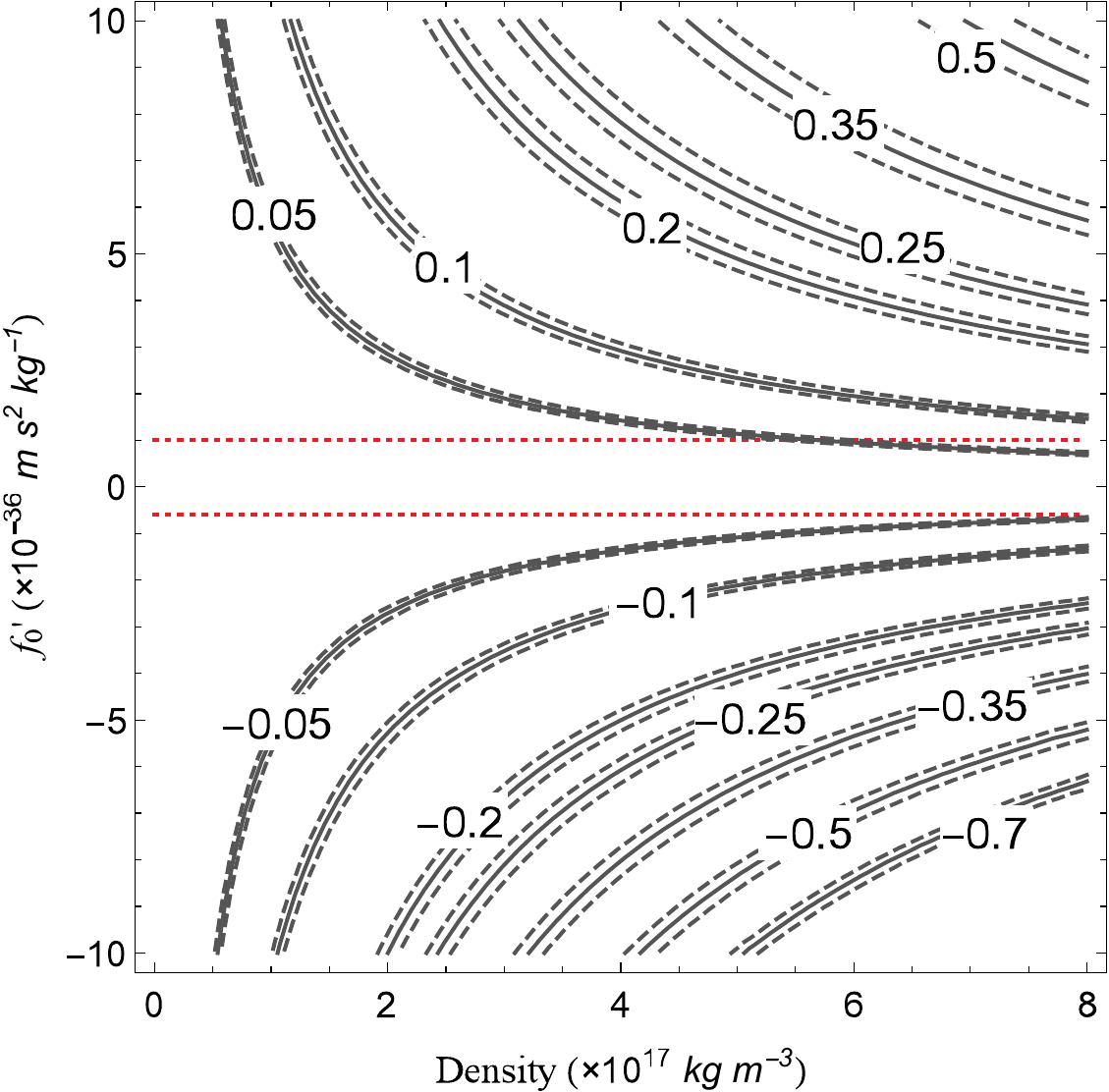}
\includegraphics[scale=0.63]{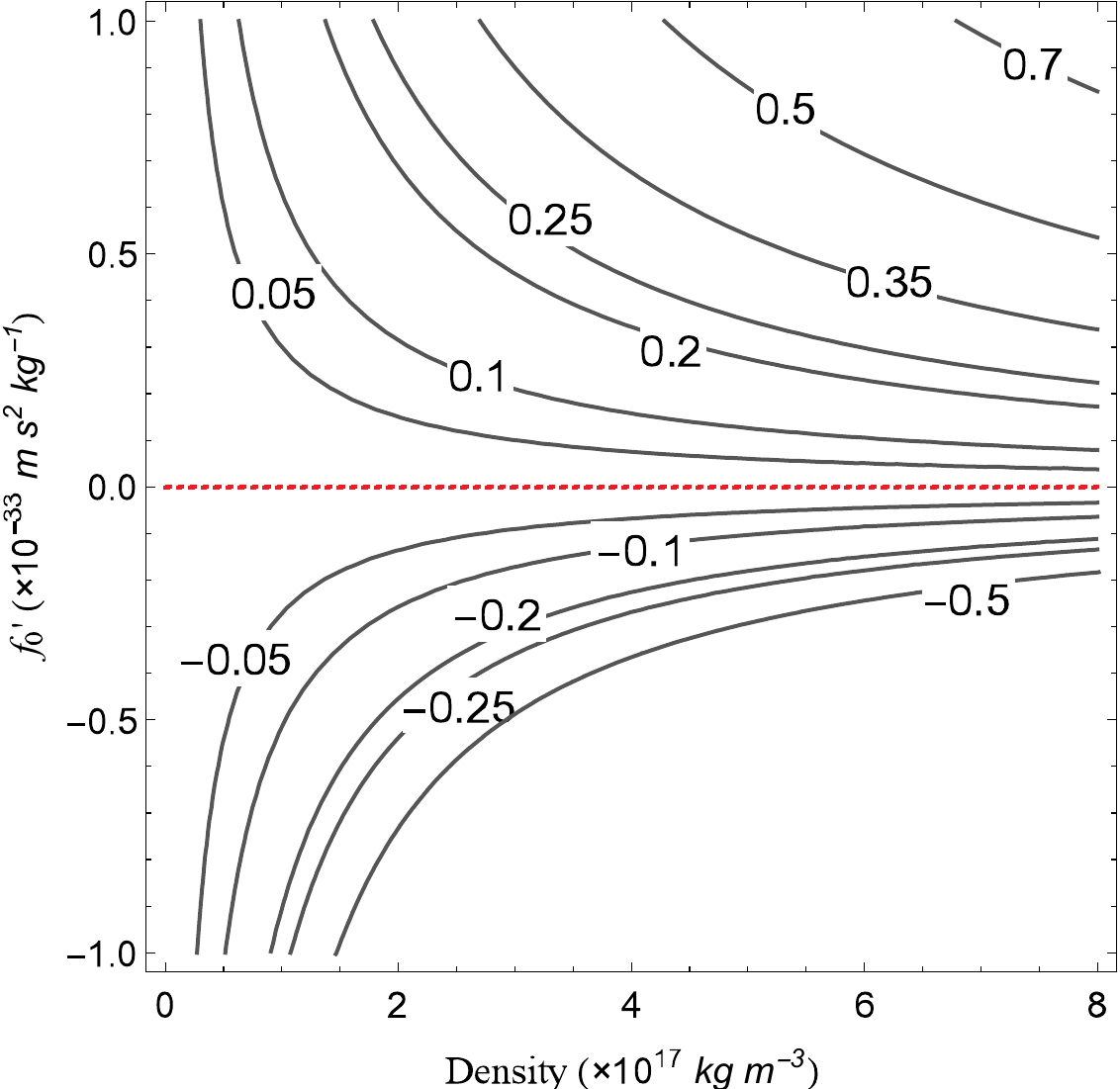}
\caption{The representation of Eq. \eqref{D} in terms of $f_0'$ and the density of the lens for three cases $\theta_{\text{s}}=0$, $\theta_{\text{s}}\ll \theta_{\text{E}}$, and $\theta_{\text{s}}\gg \theta_{\text{E}}$. The contours of equal $\mathcal{D}$ in the plane $(\rho_{\text{l}},f_0')$ are shown in the interval $-0.7\leqslant\mathcal{D} \leqslant 0.7$. In the top panel, we display the cases where $\theta_{\text{s}}=0$ and $\theta_{\text{s}}\ll \theta_{\text{E}}$ with the solid and dashed curves, respectively. Also, the upper and lower bounds on the EMSG free parameter obtained in \cite{nazari2020Constraining} are represented by the red dotted lines. The bottom panel shows the case where $\theta_{\text{s}}\gg \theta_{\text{E}}$. As $\theta^{-}_{\text{\tiny EMS}}$ does not satisfy the condition $\theta\gg \big(d_{\text{so}}/d_{\text{ls}}\big)\theta_{\text{E}}^2 $ in this case, we only study $\theta^{+}_{\text{\tiny EMS}}$ in this panel.}\label{fig1}
\end{center}
\end{figure}
%================================================================================

Inserting the EMSG light deflection \eqref{lighdef} into the above relation, we then obtain
\begin{align}\label{lenseq}
\theta^2-\theta_{\text{s}}\theta-\theta_{\text{E}}^2\big(1+2c^2f_0'\rho_{\text{l}}\big)=0,
\end{align}
in which $\rho_{\text{l}}$ is the mass density of the lens and 
\begin{align}
\theta_{\text{E}}=\Big(\frac{4GM}{c^2}\frac{d_{\text{ls}}}{d_{\text{so}}d_{\text{lo}}}\Big)^{1/2},
\end{align}
is the Einstein angle.
To find this relation, we use the fact that $b=d_{\text{lo}}\theta$ where $d_{\text{lo}}$ indicates the  angular diameter distance from the lens to the observer. We should recall that to derive this relation, it is assumed that the density of the lens is uniform. It is also worth mentioning that Eq. \eqref{lenseq} is established where the point of closest approach of the photon to the center-of-mass of the lens, $b$, is much larger than twice the Schwarzschild radius of the lens $R_{\text{Sch}}$, i.e., $b\gg 2\,R_{\text{Sch}}$. In fact, here, photons pass far away from the lens's photon sphere. It means that the solutions to the lens equation should satisfy the condition $\theta\gg \big(d_{\text{so}}/d_{\text{ls}}\big)\theta_{\text{E}}^2$ in this lensing scenario.  
This is the realm where 1\tiny PN \normalsize corrections are sufficient to describe gravitational lensing. Beyond this regime, strong-deflection lensing including higher PN terms, should be investigated.

By solving the lens equation \eqref{lenseq}, we find the position of lensed images with EMSG corrections as follows
\begin{align}\label{imagep}
\theta^{\pm}_{\text{\tiny EMS}}=\frac{1}{2}\Big(\theta_{\text{s}}\pm\sqrt{\theta_{\text{s}}^2+4\theta_{\text{E}}^2\big(1+2c^2f_0'\rho_{\text{l}}\big)}\Big).
\end{align}
As seen, like GR, there are two solutions. Furthermore, given the EMSG corrections, the position of these images deviates from the GR case.
Therefore, investigating Eq. \eqref{imagep} can provide a possible test for the EMSG theory. Let us define the following ratio
\begin{align}\label{D}
\mathcal{D}=\frac{\theta^{\pm}_{\text{\tiny EMS}}-\theta^{\pm}_{\text{\tiny GR}}}{\theta^{\pm}_{\text{\tiny GR}}},
\end{align}
where $\theta^{\pm}_{\text{\tiny GR}}=\theta^{\pm}_{\text{\tiny EMS}}(f_0'\rightarrow 0)$.  
In Fig. \ref{fig1}, we illustrate this relative deviation of the EMSG image positions and the GR ones for three cases: $i)$ where the source is completely behind the lens, i.e., $\theta_{\text{s}}=0$. $ii)$ where $\theta_{\text{s}}\ll \theta_{\text{E}}$. $iii)$ where $\theta_{\text{s}}\gg \theta_{\text{E}}$.
It should be mentioned that for the third case, $\theta^{-}_{\text{\tiny EMS}}$ does not satisfy the desired condition $\theta\gg \big(d_{\text{so}}/d_{\text{ls}}\big)\theta_{\text{E}}^2$. So, we ignore this solution and exhibit other cases in this figure.

Regarding the sign of $f_0'$, the relative deviation can be positive or negative. Here, we consider the interval $-0.7\leqslant\mathcal{D} \leqslant 0.7$. 
As mentioned earlier, we assume that the lens is a dense star with the density of the order $10^{17}\,\text{kg}\, \text{m}^{-3}$. So, the density is chosen to be in the range $10^{17}\,\text{kg}\, \text{m}^{-3}\leqslant\rho_{\text{l}}\leqslant 8\times 10^{17}\,\text{kg}\, \text{m}^{-3}$.  
In the top panel of Fig. \ref{fig1}, the relative differences for the first and second cases in the parameter space $( \rho_{\text{l}}, f_0')$ are indicated by the solid and dashed curves, respectively.
The third case is displayed by the solid curve in the bottom panel.
To visually compare our result with the previous study, the bound $-0.6\times10^{-36}\,\text{m}\,\text{s}^2\, \text{kg}^{-1}<f_0'<+10^{-36}\,\text{m}\,\text{s}^2\, \text{kg}^{-1}$ which is obtained in \cite{nazari2020Constraining}, is also added in these panels.
Since considerable relative differences only occur for a large amount of $f_0'$ in the third case, this interval of $f_0'$ turns into a single line in the bottom panel.
Both panels in this figure show that at a fixed density, the higher the free parameter $|f_0'|$, the greater the absolute value of $\mathcal{D}$. Furthermore, for a fixed EMSG parameter, $|\mathcal{D}|$ grows with increasing the lens mass density. Therefore, as expected, at a high mass density and a large $|f_0'|$, the relative difference will be considerable. It is seen that $|\mathcal{D}|>0.1$ for $|f_0'|\geqslant 10^{-36}\,\text{m}\,\text{s}^2\, \text{kg}^{-1}$ and $|f_0'|\geqslant 10^{-34}\,\text{m}\,\text{s}^2\, \text{kg}^{-1}$ in the top and bottom panels, respectively. In fact, in this region of the parameter space $(\rho_{\text{l}}, f_0')$, it is possible to detect the footprint of EMSG on the position of images. 
On the other hand, given the constraint on $f_0'$ shown by the red dotted lines in these panels and its intersection with the $\mathcal{D}=0.05$ curve, only a relative deviation up to $5\%$ can be expected for the cases $\theta_{\text{s}}=0$ and $\theta_{\text{s}}\ll \theta_{\text{E}}$; and in the case where $\theta_{\text{s}}\gg \theta_{\text{E}}$, there is no significant difference between the position of images in EMSG and GR. 

To shed light on the importance of this difference, let us estimate the EMSG correction to the Einstein angle in the microlensing regime. We consider a compact object as a gravitational lens and a star as a light source that are located a few kiloparsecs away from the Earth.  In this case, regarding Eq. \eqref{imagep}, we have $\theta_{E}^{\text{\tiny EMS}}\simeq\theta_{\text{E}}\big(1+c^2f_0'\rho_{\text{l}}\big)$. To find this relation, we keep only linear terms in $f_0'$.  Then, this yields $c^2f_0'\rho_{\text{l}}\theta_{\text{E}}$ for the EMSG correction up to the leading order. According to the sign of the EMSG parameter, the new Einstein ring can shrink or expand compared to the standard one in GR.
Now, in order to evaluate the order of magnitude of this correction, two possible cases are studied.  We suppose the supermassive black hole at the center of our Galaxy is a gravitational lens and a star at $d_{\text{ls}}=10\,\text{pc}$ is a source. In this case, the lens is located at $d_{\text{lo}}=7.9\pm 0.4\,\text{kpc}$ \cite{eisenhauer2003geometric} with $M=(3.6\pm 0.2)\times 10^{6}M_{\odot}$ \cite{ghez2005stellar}. It is assumed that the lens is a sphere with a Schwarzschild radius and its density is defined by $\rho_{\text{l}}=(3\,c^6/32\,\pi\, G^3)(1/M^2)$. Also, we set $f_0'=10^{-36}\,\text{m}\,\text{s}^2\, \text{kg}^{-1}$. Using this definition and ignoring the uncertainty in mass and distances, we find that the EMSG correction to the Einstein angle for this lens is $8.7\times 10^{-9} \mu\text{ac}$ (micro-arcseconds). Then, for this supermassive black hole, the EMSG contribution to the position of images is insignificant and it is exceedingly faint to be detected. On the other hand, one can show that in the same lensing scenario with a lighter black hole, e.g., $M=50\,M_{\odot}$, the order of magnitude of the EMSG correction is about $0.2\,\mu\text{ac}$.   
For the next case, we choose another possible system in which the stellar lens and source are located in the Galactic halo and Magellanic Clouds at $d_{\text{lo}}=20\,\text{kpc}$ and $d_{\text{so}}=50\,\text{kpc}$, respectively. We also assume that the lens is a neutron star with $\rho_{\text{l}}=10^{17}\,\text{kg}\,\text{m}^{-3}$ and set $f_0'=10^{-36}\,\text{m}\,\text{s}^2\, \text{kg}^{-1}$. In this case, the EMSG correction to the Einstein angle is of the order $5.3\,\mu\text{ac}$. 
Moreover, for $f_0'=10^{-37}\,\text{m}\,\text{s}^2\, \text{kg}^{-1}$, which is one order of magnitude smaller than the upper limit of the EMSG free parameter obtained in \cite{nazari2020Constraining,akarsu2018constraint}, one can find that this correction would be of the order $0.5\,\mu\text{ac}$.  
From these interesting astrophysical cases, we deduce that the EMSG correction can be of the order $(1-0.1)\,\mu\text{ac}$. Although this change in the image position is small, it is expected to be measurable by future high-resolution missions such as MAXIM \cite{cash2000laboratory,cash2004maxim}. Indeed, this may provide a direct observational test for this theory in the weak-field limit and improve the constraint on the EMSG free parameter, its upper limit, by at least one order of magnitude.
As the deviation for $f_0'=-10^{-38}\,\text{m}\,\text{s}^2\, \text{kg}^{-1}$, which is one order of magnitude bigger than the lower limit of the EMSG free parameter, is of the order $0.05\,\mu\text{ac}$, constraining the lower limit of $f_0'$ is beyond the sensitivity of the planned microarcsecond-resolution missions.

From the observational point of view, the modification of the Einstein angle could affect the measurement of some quantities. To clarify this, we study the characteristic timescale of a microlensing event given by $t_{\text{E}}=R_{\text{E}}/v_{\perp}$. Here, $R_{\text{E}}=d_{\text{lo}}\theta_{\text{E}}$ is the Einstein radius and $v_{\perp}$ is the transverse velocity of the lens relative to the line of sight. 
This timescale provides a tool to estimate the lens mass.
Obviously, the difference of the Einstein angle due to the extra EMSG potential gives rise to the change of the Einstein time $t_{\text{E}}$. Here, we assume that the distances and the lens transverse velocity are known. So, this change means that the lens mass is different from what is obtained in GR. In fact, by setting $t_{\text{E}}^{\text{\tiny EMS}}=t_{\text{E}}^{\text{obs}}=t_{\text{E}}^{\text{\tiny GR}}$, one can arrive at $M_{\text{\tiny EMS}}\simeq t_{\text{E},\text{obs}}^{2}\,x\big(1-2c^2f_0'\rho_{\text{l}}\big)$, whereas we have $M_{\text{\tiny GR}}= t_{\text{E},\text{obs}}^{2}\,x$ in GR. Here, $x\equiv(c^2/4G)\big(d_{\text{so}}/(d_{\text{lo}}d_{\text{ls}})\big)v^2_{\perp}$. 
Therefore, the lens mass may be overestimated or underestimated utilizing the classical/GR method.
The EMSG correction to the Einstein time also reveals that depending on the density of the lens and the free parameter of the theory, the mass of the compact object, for instance, the neutron star, in the EMSG theory could be smaller or larger than the standard case inferred from GR. This fact is consistent with the results of \cite{nari2018compact}. Nonetheless, we emphasize that a more complete analysis should be adopted when the distances in this theory are properly studied.

%================================================================================
\begin{figure}
\begin{center}
\centering
\includegraphics[scale=0.90]{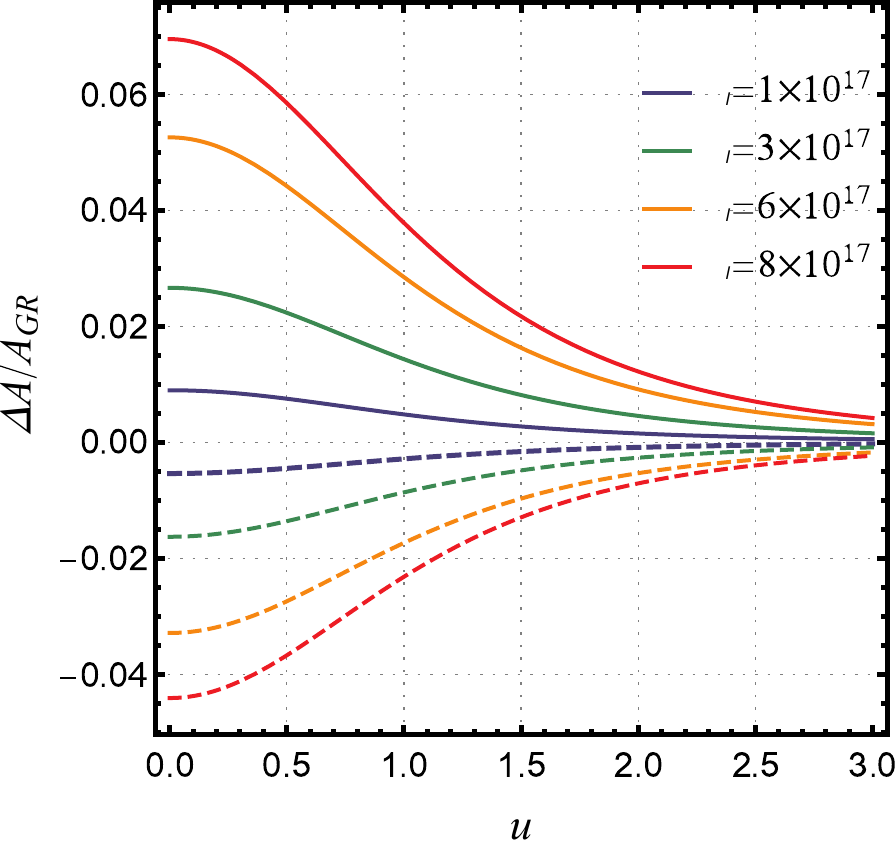}
\caption{The relative difference between $A_{\text{\tiny EMS}}$ and $A_{\text{\tiny GR}}$ in terms of $u$ for different values of the lens density. Here, the solid and dashed curves show the cases with $f_0'=10^{-36}\,\text{m}\,\text{s}^2\, \text{kg}^{-1}$ and $f_0'=-0.6\times10^{-36}\,\text{m}\,\text{s}^2\, \text{kg}^{-1}$, respectively. These values are the upper and lower limits on the EMSG free parameter which are obtained in \cite{nazari2020Constraining}.}\label{fig2}
\end{center}
\end{figure}
%================================================================================

Other interesting quantity studied in microlensing is the magnification of the primary and secondary images which is defined as
\begin{align}
&A^{\pm}=\Big|\frac{\theta^{\pm}}{\theta_{\text{s}}}\frac{d\theta^{\pm}}{d \theta_{\text{s}}}\Big|,
\end{align}
for a steady and very small source \cite{paczynski1996gravitational}. Given this relation and Eq. \eqref{imagep}, for a similar source, we define the magnification in the EMSG theory as
\begin{align}
\nonumber
&A^{\pm}_{\text{\tiny EMS}}=\Big|\frac{\theta^{\pm}_{\text{\tiny EMS}}}{\theta_{\text{s}}}\frac{d\theta^{\pm}_{\text{\tiny EMS}}}{d \theta_{\text{s}}}\Big|\\
&=\frac{1}{u\sqrt{u^2+8c^2f_0'\rho_{\text{l}}+4}}\Big(\frac{u^2}{2}+2c^2f_0'\rho_{\text{l}}+1\Big)\pm\frac{1}{2},
\end{align}
where $u\equiv\theta_{\text{s}}/\theta_{\text{E}}$. 
Since the separation of images is too small in a microlensing event, the images cannot be detected individually and only the total magnification of a source can be observed. So, we focus our attention on the total magnification. Summing $A^{+}_{\text{\tiny EMS}}$ and $A^{-}_{\text{\tiny EMS}}$, we arrive at
\begin{align}\label{totmag}
A_{\text{\tiny EMS}}
=\frac{1}{u\sqrt{u^2+8c^2f_0'\rho_{\text{l}}+4}}\Big(u^2+4c^2f_0'\rho_{\text{l}}+2\Big),
\end{align}
for the total magnification in the framework of EMSG.
For $f_0'=0$, this relation reduces to the GR one, cf. Eq. (11) in \cite{paczynski1996gravitational}. It shows the consistency of the EMSG theory.

%================================================================================
\begin{figure}
\begin{center}
\centering
\includegraphics[scale=0.80]{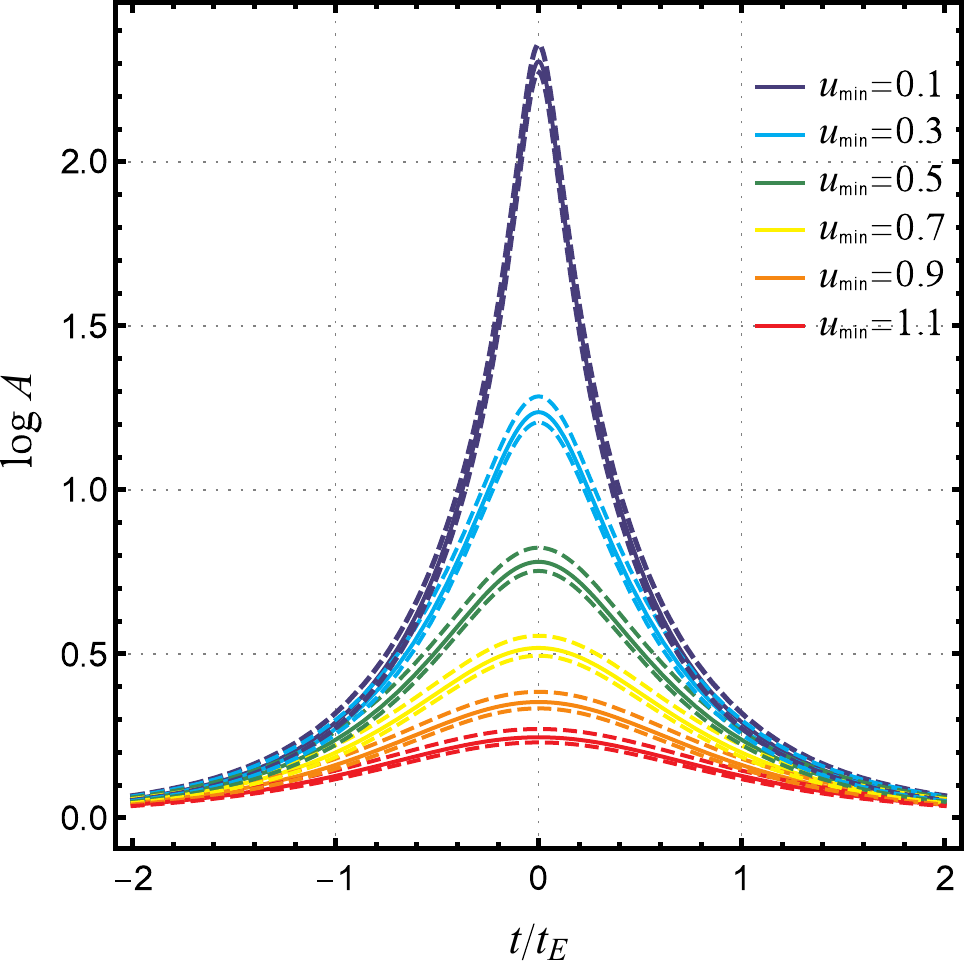}
\caption{The variation of the combined magnification with the dimensionless time $t/t_{\text{E}}$ for $u_{\text{min}}=0.1, 0.3, 0.5, 0.7, 0.9, 1.1$.  The solid and dashed curves exhibit the GR and EMSG cases, respectively. Also, the dashed curves above and below the solid ones stand for the EMSG case with $f_0'=10^{-36}\,\text{m}\,\text{s}^2\, \text{kg}^{-1}$ and $f_0'=-0.6\times10^{-36}\,\text{m}\,\text{s}^2\, \text{kg}^{-1}$, respectively. Here, we set $\rho_{\text{l}}=6\times10^{17}\,\text{kg}\,\text{m}^{-3}$. }\label{fig3}
\end{center}
\end{figure}
%================================================================================

In order to reveal the influence of the EMSG terms on the total magnification, we exhibit the relative difference $(A_{\text{\tiny EMS}}-A_{\text{\tiny GR}})/A_{\text{\tiny GR}}$ in terms of $u$ in Fig. \ref{fig2}. In this figure, the solid and dashed curves respectively belong to the upper and lower bounds of the EMSG parameter represented in \cite{nazari2020Constraining}. This deviation is also studied in the density range $10^{17}\,\text{kg}\,\text{m}^{-3}\leqslant\rho\leqslant8\times10^{17}\,\text{kg}\,\text{m}^{-3}$. As seen, the deviation grows with decreasing $u$ as well as increasing $\rho_{\text{l}}$. It means that in the situation where the source is almost behind the dense lens, the EMSG correction is more effective, and as a result, the combined light intensity could be different from the GR one.  Given the solid curves, all of which take positive values in this figure, for $f_0'=10^{-36}\,\text{m}\,\text{s}^2\, \text{kg}^{-1}$, the images in EMSG are brighter than those in GR. In the best case, the EMSG image is about $7\%$ brighter. The situation is quite the opposite for $f_0'=-0.6\times10^{-36}\,\text{m}\,\text{s}^2\, \text{kg}^{-1}$, and the EMSG image is dimmer than its GR counterpart. See the dashed curves for which $A_{\text{\tiny EMS}}<A_{\text{\tiny GR}}$. Here, in the best case, the EMSG image is about $4.5\%$ dimmer. Therefore, the light source in the EMSG theory would be microlensed differently compared to GR; and for the positive/negative value of $f_0'$, the EMSG image is brighter/fainter than the GR image. This fact is also in agreement with the results inferred from Fig \ref{fig1}.

%================================================================================
\begin{figure}
\begin{center}
\centering
\includegraphics[scale=0.85]{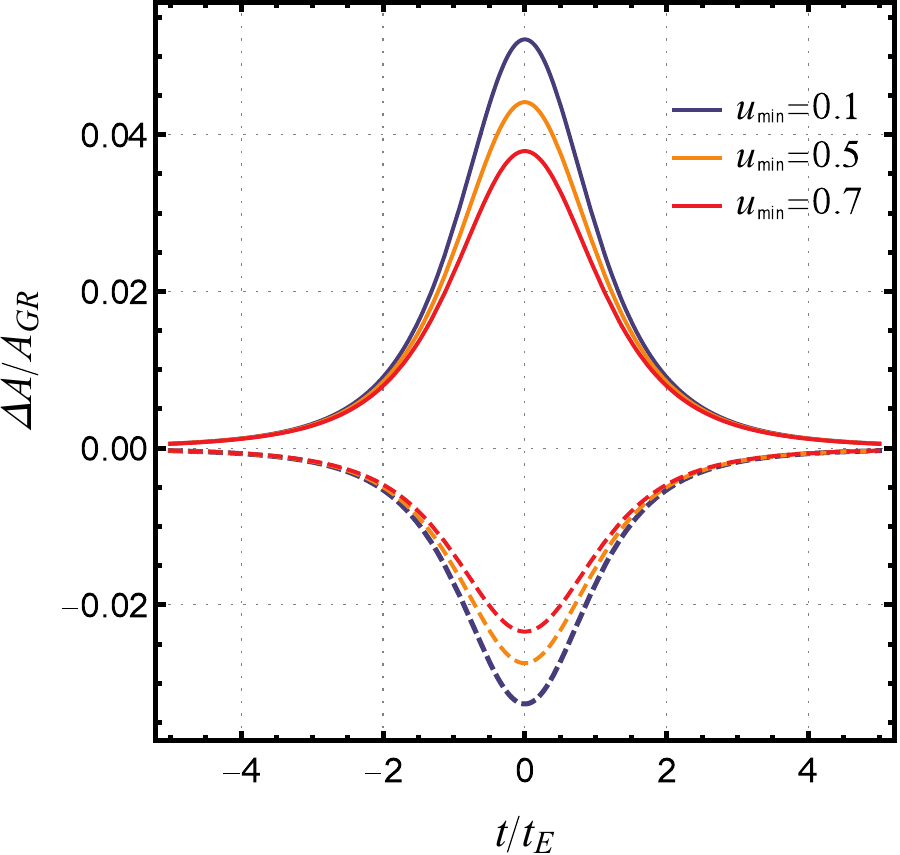}
\caption{The relative difference between $A_{\text{\tiny EMS}}$ and $A_{\text{\tiny GR}}$ in terms of the dimensionless time $t/t_{\text{E}}$ for different values of $u_{\text{min}}$. 
The solid and dashed curves show the cases with $f_0'=10^{-36}\,\text{m}\,\text{s}^2\, \text{kg}^{-1}$ and $f_0'=-0.6\times10^{-36}\,\text{m}\,\text{s}^2\, \text{kg}^{-1}$, respectively. Here, we assume that $\rho_{\text{l}}=6\times10^{17}\,\text{kg}\,\text{m}^{-3}$. }\label{fig4}
\end{center}
\end{figure}
%================================================================================ 

In the framework of microlensing, it is also interesting to study the shape of the light curve versus the time it takes for the lens to move relative to the source. Here, we are interested in studying the total magnification in terms of the duration of a microlensing phenomenon in the EMSG theory. To do so, we assume that the lens has a uniform motion. It is shown that in this case, the position of the source in terms of time is given by \cite{paczynski1986gravitational,paczynski1996gravitational}
\begin{align}
u=\Big[u_{\text{min}}^2+\Big(\frac{v_{\perp}(t-t_0)}{R_{\text{E}}}\Big)^2\Big]^{1/2},
\end{align}
where $u_{\text{min}}=u(t=t_{\text{0}})$ is the dimensionless impact parameter and $t_0$ is the time of closest approach to the lens. In the following calculations, without losing generality, we choose $t_0=0$. 
Substituting this relation back within Eq. \eqref{totmag}, we study the change of the EMSG combined magnification in terms of time. Our results are summarized in Fig. \ref{fig3}. In this figure, for different values of $u_{\text{min}}$, $\log A$ is studied. Here, the solid and dashed curves exhibit the GR and EMSG cases, respectively.  The dashed curves above and below the solid ones belong to the EMSG case with $f_0'=10^{-36}\,\text{m}\,\text{s}^2\, \text{kg}^{-1}$ and $f_0'=-0.6\times10^{-36}\,\text{m}\,\text{s}^2\, \text{kg}^{-1}$, respectively. This figure reveals that except for a small deviation, the overall behavior of the light curves in EMSG is similar to that in GR. Indeed, the smaller the dimensionless impact parameter $u_{\text{min}}$, the brighter the lensed image. This fact significantly occurs at $t=t_0=0$. To indicate when the deviation between the standard and EMSG cases is considerable, we also study $(A_{\text{\tiny EMS}}-A_{\text{\tiny GR}})/A_{\text{\tiny GR}}$ in terms of time for $u_{\text{min}}<1$ in Fig. \ref{fig4}. In this figure, the solid and dashed curves belong to the upper and lower bounds of the EMSG parameter,  respectively. It is seen that in the closest approach, which occurs at $t=t_0=0$, the absolute value of the relative difference will be maximum. Also, similar to Fig. \ref{fig2}, this deviation increases with decreasing $u_{\text{min}}$.

\section{Summary and Conclusion}\label{Summary and Conclusion} 

In this work, we have studied the behavior of light rays in the weak-field limit of EMSG. 
The PN metric of an EMSG fluid has been derived.  This is the main material needed to obtain EMSG corrections to the propagation of light rays. We have utilized the modern approach to the PN theory.
It has been shown that in addition to the Newtonian potential, the EMSG potential, $U_{\text{\tiny EMS}}$, affects both longitudinal and transverse terms in the trajectory of photons. As a consequence, in this theory, photons do not behave as predicted by GR.
To understand if there is a detectable signature of the EMSG theory in the motion of light, we have studied light deflection, Shapiro time delay, and gravitational microlensing.
As a first step, it has been assumed that the deflector is compact and spherically symmetric. In fact, because the quadratic term $\rho^2$ appears in the Poisson integral of $U_{\text{\tiny EMS}}$, the point-mass description cannot be used here.
For the sake of simplification, throughout this paper, we have also considered that the density of the body is uniform. Otherwise, due to the appearance of the time-dependent EMSG term in the definition of light deflection, we would encounter more complicated calculations. Choosing the constant density body allows us to easily evaluate and understand the EMSG correction to the motion of light.

It has been shown that the EMSG contribution to light deflection and Shapiro time delay is similar and in both cases, the same expression $\big(1+2c^2f_0'\rho\big)$ is added to the classical one.
It means that in addition to the free parameter of the theory, $f_0'$, the magnitude of the EMSG correction strongly depends on the density of the deflector.  
Therefore, for the dense system where gravity experiments are carried out, the EMSG effects can be significant or even detectable. 
Regarding the results obtained in \cite{bertotti2003test,shapiro2004measurement}, we have restricted the free parameter of the theory and shown that it should lie within the range $-4.0\times 10^{-27}\, \text{m}\,\text{s}^2\,\text{kg}^{-1} < f_0' < 8.7\times 10^{-26}\,\text{m}\,\text{s}^2\,\text{kg}^{-1}$ to justify the solar system tests such as light bending and Shapiro time delay.
Since there is no inconsistency with this interval and those derived in \cite{nazari2020Constraining,akarsu2018constraint}, we claim that the EMSG theory passes these solar system tests with flying colors. However, more accurate measurements are needed to find a tighter bound on the free parameter of the theory and to observe its possible signature in the solar system framework. In other words, EMSG and GR cannot be distinguished only using these classical tests of gravity with the current accuracy.
The main reason that the interval obtained from these tests is not more limited than what is introduced in the precedent studies is that the gravitational system considered here is the Sun, which has a very low density. 

To examine possible detectable signatures of EMSG in light curves, we have next focused our attention on a source microlensed by a dense object like neutron stars. 
Notably, there are two images in the EMSG gravitational lensing scenario whose positions deviate from those in GR. It has been shown that given the sign of the EMSG parameter, the new Einstein ring can shrink or expand compared to the standard one. 
Two interesting and possible astrophysical systems have been applied to estimate the EMSG correction to the Einstein angle in the microlensing regime. For these cases, we have predicted that this correction would be as large as $(1-0.1)\,\mu\text{ac}$ which could be detected by future high-resolution missions such as MAXIM \cite{cash2000laboratory,cash2004maxim}. This advance in measurement may not only distinguish this modified theory of gravity from GR, but can place a stronger observational constraint on the free parameter of the theory compared to the previous studies in this context. It may improve the upper limit of $f_0'$ by at least one order of magnitude.

We have then investigated the total magnification and the shape of light curves in the EMSG theory. It is revealed that except for a small shift, the overall behavior of the EMSG light curves in terms of time is similar to that in GR.
It also turns out that where the light source is almost behind a dense lens, the EMSG correction is more efficient, and the combined light of the lensed images is different from the GR case. 
Remarkably, compared to GR, for the positive $f_0'$, the EMSG image becomes brighter, while for the negative $f_0'$, the EMSG image becomes dimmer.   
In the best case, this deviation from GR is less than $10\%$. However, it is possible that in the case of the strong magnification events with small measurement errors, the signature of EMSG theory due to the extra gravitational potential can be detected.

To sum up, the solar system tests such as light bending and Shapiro time delay do not rule out this theory. In the gravitational microlensing scenario, a dense compact lens like neutron stars allows us to distinguish this modified gravity from GR. It is possible that future high-resolution missions could provide an observational test for the EMSG theory in the weak-field limit.

\section*{acknowledgments}

I would like to thank Mahmood Roshan for reading this paper and for his useful suggestions. This work is supported by Ferdowsi University of Mashhad.

\bibliographystyle{apsrev4-1}
\bibliography{short,PN_EMSG}

\newpage

\appendix

\section{Near-zone and wave-zone solutions of the wave equation}\label{app_1}

In this appendix, we derive the approximate solution of the highly non-linear wave equation \eqref{wave_eq} where the field point is situated within the near zone. Utilizing this solution and applying expansions \eqref{g00}-\eqref{gjk}, one can systematically construct the near-zone spacetime metric of a system to an adequate degree of accuracy. To find the solution to the equation \eqref{wave_eq}, we take the advantage of the iterative procedure introduced in \cite{poisson2014gravity}. The main idea behind this method is to approximately linearize this equation. Then using the retarded Green's function, one can integrate the linearized wave equation \footnote{ This technique is comprehensively introduced in chapter 6 of \cite{poisson2014gravity}. As the mathematical form of the EMSG field equations in landau-Lifshitz formalism is similar to the GR one, we use the general retarded solutions to the wave equation introduced in this reference.}. To do so, the source term of each iteration of the wave equation, i.e., $\tau^{\mu\nu}_{\text{eff}}$, is built in the previous step. In fact, in this manner, Eq. \eqref{wave_eq} is no longer non-linear in terms of $h^{\mu\nu}$ and the wave equation can be integrated straightforwardly in each iterative step. In the following, each step is indicated with the index $(n)$ where $n$ shows the number of the iteration of the wave equation. We should then solve $\square h^{\mu\nu}_{(n)}=-2k\tau^{\mu\nu}_{\text{eff}(n-1)}$ in every step.

It should also be mentioned that the method of extracting the PM and PN approximations generally is similar and the PN limit is in fact embedded within the PM approximation. Therefore, some parts of our calculations, more specifically the first iteration, inevitably overlap with those of \cite{nazari2020Constraining}. However, for the sake of completeness, we discuss these parts in detail here.

Before getting our hands dirty with the iteration procedure, let us collect all general forms of the solutions to the wave equation we need during our calculations. Regarding the position of the field and source points of the wave equation, these general solutions are categorized. We rewrite those in which the field point is located in the near-zone region of spacetime.
According to the naming rule in \cite{poisson2014gravity}, the solution whose source point is located in the near (wave) zone is indicated by the index $\mathcal{N}$ ($\mathcal{W}$) and called the near-zone (wave-zone) solution.

The first general solution to the wave equation is the near-zone solution given by 
\begin{align}
 \nonumber
 h_{\mathcal{N}}^{\mu\nu}{(t,\boldsymbol{x})}=&\frac{k}{2}\sum_{l=0}^{\infty}\frac{(-1)^l}{l! c^l}\\\label{hNear}
&\times\Big(\frac{\partial}{\partial t}\Big)^l
\int_{\mathcal{M}}\tau^{\mu\nu}_{\text{eff}}{(t,\boldsymbol{x}')}\rvert{\boldsymbol{x}-\boldsymbol{x}'}\rvert^{l-1} d^3x',
\end{align}
in which the source, $\boldsymbol{x}'$, and field, $\boldsymbol{x}$, points both are situated in the near zone.
If we consider $\mathcal{R}$ to be the boundary between the near and wave zones, we have $r'=\lvert\boldsymbol{x}'\lvert<\mathcal{R}$ and $r=\lvert\boldsymbol{x}\lvert<\mathcal{R}$ in the above relation. 
Here, $\mathcal{M}$ is a three-dimensional sphere with the radius $\mathcal{R}$ representing the near-zone region. In the following, we show its boundary with $\partial \mathcal{M}$.

The second one is the wave-zone solution where the field and the source points are situated in the near and the wave zones, respectively. This solution is introduced as follows:
\begin{align}
\nonumber
& h^{\mu\nu}_{\mathcal{W}}{(t,\boldsymbol{x})}=\frac{k}{2}\frac{n^{<j_1j_2\cdots j_l>}}{r}\bigg\lbrace\int_{\mathcal{R}-r}^{\mathcal{R}}f^{\mu\nu}(\tau-2s/c)A(s,r)ds\\\label{eq8}
&+\int_{\mathcal{R}}^{\infty}f^{\mu\nu}(\tau-2s/c)B(s,r)ds\bigg\rbrace,
\end{align}
in which $A(s,r)=\int_{\mathcal{R}}^{r+s}P_{l}(\zeta)p^{1-n}dp$ and  $B(s,r)=\int_{s}^{r+s}P_{l}(\zeta)p^{1-n}dp$ where $P_{l}(\zeta)$ is a Legendre polynomial. Here, $\zeta=(r+2s)/r-2s(r+s)/(rp)$, and $n^{<j_1j_2\cdots j_l>}$ is an angular symmetric trace-free tensor. cf. Eq. (1.154) of \cite{poisson2014gravity}. Moreover, for this solution, the source term of the wave equation is written as
\begin{align}\label{eq7}
\tau^{\mu\nu}_{\text{eff}}=\frac{1}{4\pi}\frac{f^{\mu\nu}(\tau)}{r^n}n^{<j_1j_2\cdots j_l>}.
\end{align}
Here, $\tau=t-r/c$ is the retarded time.

\subsection{Zeroth and First iterations}

We start the iterative method with the zeroth step.  
At this stage, the spacetime is described by the Minkowski metric, i.e.,  $g^{\mu\nu}_{(0)}=\eta^{\mu\nu}$ and $\sqrt{-g^{(0)}}=1$. Then, we have $h^{\mu\nu}_{(0)}=0$.
Regarding this, we construct the components of the effective energy-momentum tensors $T^{\mu\nu}_{\text{eff(0)}}$. As mentioned, we choose a perfect fluid to describe the matter part of the system.  Considering the definition of $T^{\mu\nu}_{\text{eff}}$, we arrive at

\begin{subequations}
\begin{align}
\nonumber
& c^{-2}T^{00}_{\text{eff(0)}}=\rho^*+{\rho^*}^2c^2 f_0'\Big[1+\frac{1}{c^2}\big(v^2+2\Pi+\frac{8p}{\rho^*}\big)\Big]\\\label{T00(0)EMS}
&~~~~~~~~~~~~~~~+O(c^{-2}),\\
\nonumber
& c^{-1}T^{0j}_{\text{eff(0)}}=\rho^*v^j+4{\rho^*}^2c^2v^jf_0'\Big[\frac{1}{2}+\frac{1}{c^2}\big(\Pi+\frac{2p}{\rho^*}\big)\Big]\\\label{T0j(0)EMS}
&~~~~~~~~~~~~~~~+O(c^{-2}),\\
\nonumber
&
 T^{jk}_{\text{eff(0)}}={\rho^*}^2c^4f_0'\Big[\delta^{jk}-\frac{1}{c^2}\Big(\big(v^2-2\Pi\big)\delta^{jk}\\\label{T(0)EMSG}
&~~~~~~~~~-2v^jv^k\Big)\Big]+O(1),
\end{align}
\end{subequations}
where $\rho^*=\sqrt{-g}\gamma \rho$ is the rescaled mass density and $\Pi=\epsilon/\rho^*$. In the zeroth iteration, one can show that $\gamma=1+\frac{1}{2}\frac{v^2}{c^2}+O(c^{-4})$ and  $\rho^*=\big(1+\frac{1}{2}\frac{v^2}{c^2}+O(c^{-4})\big)\rho$. It should be mentioned to find Eqs. \eqref{T00(0)EMS}-\eqref{T(0)EMSG}, we apply the normalization condition $g_{\mu\nu}u^\mu u^\nu=-c^2$.
To completely construct the source term of the wave equation, we also need to build the landau-Lifshitz and harmonic pseudotensors at this stage. The general form of these pseudotensors is defined by
\begin{align}
\nonumber
&(-g)t_{\text{LL}}^{\alpha\beta}=\frac{1}{2k}\bigg\lbrace\frac{1}{2}\eta^{\alpha\beta}\eta_{\lambda\mu}\partial_{\rho}h^{\lambda\nu}\partial_{\nu}h^{\mu\rho}-\eta^{\alpha\lambda}\eta_{\mu\nu}\partial_{\rho}h^{\beta\nu}\\\nonumber
&\times\partial_{\lambda}h^{\mu\rho}-\eta^{\beta\lambda}\eta_{\mu\nu}\partial_{\rho}h^{\alpha\nu}\partial_{\lambda}h^{\mu\rho}+\eta_{\lambda\mu}\eta^{\nu\rho}\partial_{\nu}h^{\alpha\lambda}\partial_{\rho}h^{\beta\mu}\\\label{tLL}
&+\frac{1}{8}\big(2\eta^{\alpha\lambda}\eta^{\beta\mu}-\eta^{\alpha\beta}\eta^{\lambda\mu}\big)\big(2\eta_{\nu\rho}\eta_{\sigma\tau}-\eta_{\rho\sigma}\eta_{\nu\tau}\big)\partial_{\lambda}h^{\nu\tau}\partial_{\mu}h^{\rho\sigma}\bigg\rbrace,
\end{align}
as well as
\begin{align}\label{tH}
(-g)t_{\text{H}}^{\alpha\beta}=\frac{1}{2k}\big(\partial_{\mu}h^{\alpha\nu}\partial_{\nu}h^{\beta\mu}-h^{\mu\nu}\partial_{\mu\nu}h^{\alpha\beta}\big),
\end{align}
where the harmonic gauge condition is imposed.
Substituting $h^{\mu\nu}_{(0)}=0$ in the above relations, one finds that $t_{\text{LL}(0)}^{\mu\nu}=0=t_{\text{H}(0)}^{\mu\nu}$. 
Therefore, Eqs. \eqref{T00(0)EMS}-\eqref{T(0)EMSG} would be the source terms of the wave equation in the next iteration. It will be shown that the extra EMSG terms in these relations induce different gravitational potentials in the next steps. So, this is the starting point of departure from the PN limit of GR. 

Now, we solve the wave equation $\square h^{\mu\nu}_{(1)}=-2k\tau^{\mu\nu}_{\text{eff}(0)}$ to find  $h_{(1)}^{\mu\nu}=h_{\mathcal{N}(1)}^{\mu\nu}+h_{\mathcal{W}(1)}^{\mu\nu}$.
We first focus our attention on the near-zone solution $h_{\mathcal{N}(1)}^{\mu\nu}$.
We arrive at
\begin{subequations}
\begin{align}
\label{h00_N1}
& h_{\mathcal{N}(1)}^{00}=\frac{4}{c^2}U+4f_0'U_{\text{\tiny EMS}}+O(c^{-3}),\\
\label{h0j_N1}
& h^{0j}_{\mathcal{N}(1)}=\frac{4}{c^3}U^{j}+\frac{8}{c}f_0'U^j_{\text{\tiny EMS}}+O(c^{-4}),\\
\label{hjk_N1}
& h^{jk}_{\mathcal{N}(1)}= 4f_0'\delta^{jk}\big( U_{\text{\tiny EMS}}-\frac{G}{c}\frac{d}{dt}\mathfrak{M}\big)+O(c^{-4}),
\end{align}
\end{subequations}
after inserting Eqs. \eqref{T00(0)EMS}-\eqref{T(0)EMSG} into integral \eqref{hNear}. Here, the gravitational potentials $U$, $U^j$, $U_{\text{\tiny EMS}}$, and $U^j_{\text{\tiny EMS}}$ are respectively given by  
\begin{subequations}
\begin{align}
\label{U}
& U=G\int_{\mathcal{M}}\frac{{\rho^*}'}{\rvert{\boldsymbol{x}-\boldsymbol{x}'}\rvert}d^3x',\\
& U^j=G\int_{\mathcal{M}}\frac{{\rho^*}'v'^j}{\rvert{\boldsymbol{x}-\boldsymbol{x}'}\rvert}d^3x',\\
\label{U_EMSG}
& U_{\text{\tiny EMS}}=G\int_{\mathcal{M}}\frac{{{\rho^*}'}^2}{\rvert\boldsymbol{x}-\boldsymbol{x}'\rvert}d^3x'\\
\label{UjEMSG}
& U^j_{\text{\tiny EMS}}=G\int_{\mathcal{M}}\frac{{{\rho^*}'}^2v'^j}{\rvert{\boldsymbol{x}-\boldsymbol{x}'}\rvert}d^3x'.
\end{align}
\end{subequations}
We also introduce a new parameter $\mathfrak{M}$ in EMSG as 
\begin{align}\label{eq1}
& \mathfrak{M}=\int_{\mathcal{M}}{{\rho^*}'}^2d^3x'.
\end{align}

The next case is the wave-zone solution, $h_{\mathcal{W}(1)}^{\alpha\beta}$.
Regarding the position of the source point as well as the slow-motion condition and weak-field limit, in this case, the only source term of the wave equation can be $t_{\text{LL}}^{\mu\nu}$ and $t_{\text{H}}^{\mu\nu}$. On the other hand, it is shown that these terms vanish in the zeroth step. Therefore, $h_{\mathcal{W}(1)}^{\alpha\beta}=0$ and consequently $h_{(1)}^{\alpha\beta}=h_{\mathcal{N}(1)}^{\alpha\beta}$.    
 
We can now construct the near-zone metric in the first iteration. After substituting Eqs. \eqref{h00_N1}-\eqref{hjk_N1} within Eqs. \eqref{g00}-\eqref{g} and some simplifications, we obtain

\begin{subequations}
\begin{align}
\label{g00_1}
&g_{00}^{(1)}=-1+\frac{2U}{c^2}+8f_0'U_{\text{\tiny EMS}}+O(c^{-3}),\\
\label{g0j_1}
&g_{0j}^{(1)}=-\frac{4U^j}{c^3}-\frac{8}{c}f_0'U^j_{\text{\tiny EMS}}+O(c^{-4}),\\
\label{gij_1}
&g_{jk}^{(1)}=\Big(1+\frac{2U}{c^2}\Big)\delta_{jk}+O(c^{-4}),\\
&(-g_{(1)})=1+\frac{4U}{c^2}-8f_0'U_{\text{\tiny EMS}}+O(c^{-3}).
\end{align}
\end{subequations}
Given the PN corrections of the metric components, one can easily grasp that this metric does not have enough information to describe a relativistic system up to the 1\tiny PN \normalsize order. Therefore, we continue the iterative procedure until the desired relativistic corrections are achieved. As the final point of this part, let us mention that the strange order $c^{-3}$ in the determinant and time-time component of the metric is completely constructed from an EMSG term. However, since we are in the first iterated step, this term cannot be trusted to check the time-reversal invariance of solutions.
By completing the second iteration, we find that this odd order would not exist in the PN expansion of this component of the metric.

As claimed, each order $f_0'$ is treated as a PN correction. Considering the above PN expansions of the metric components and comparing terms with each other, for instance, the first and third terms in the time-time component, one can conclude that $f_0'$ must be very small and at most of the order of $c^{-2}$. So our previous assumption made in Sec. \ref{The strategy of calculations} is justifiable.

\subsection{Second iteration}

By using the metric built in the previous iterated step, we can now derive the main materials to construct the PN expansion of $h_{(2)}^{\mu\nu}=h_{\mathcal{N}(2)}^{\mu\nu}+h_{\mathcal{W}(2)}^{\mu\nu}$. 
We launch our calculations by constructing the source term of $h_{\mathcal{N}(2)}^{\mu\nu}$.  
It should be mentioned that to find the desired PN order for $h_{(2)}^{\mu\nu}$ and consequently for $g^{\mu\nu}_{(2)}$, 
it is necessary to build $O(1)$ for $\tau^{00}_{\text{ eff(1)}}$, order $O(c^{-1})$ for $\tau^{0j}_{\text{eff(1)}}$, and order $O(1)$ for $\tau^{jk}_{\text{eff(1)}}$. In the following, each portion of the effective energy-momentum pseudotensor is derived in detail.

Utilizing the components of $g^{\mu\nu}_{(1)}$, we find the contravariant components of the standard energy-momentum tensor of the perfect fluid as 
\begin{subequations}
\begin{align}
\label{T00_1}
& T^{00}_{(1)}=\rho^*c^2\Big[1+8f_0'U_{\text{\tiny EMS}}+\frac{1}{c^2}\big(\frac{1}{2}v^2+\Pi-U\big)\Big]+O(c^{-1}),\\\nonumber
& T^{0j}_{(1)}=\rho^*v^jc
\Big[1+8f_0'U_{\text{\tiny EMS}}+\frac{1}{c^2}\big(\frac{1}{2}v^2+\Pi+\frac{p}{\rho^*}-U\big)\Big]\\
&+O(c^{-2}),\\
\label{T^jk_1}
& T^{jk}_{(1)}=p\delta^{jk}+\rho^*v^jv^k+O(c^{-2}).
\end{align}
\end{subequations} 
We use the fact that 
$\gamma^{(1)}=1+4f_0'U_{\text{\tiny EMS}}+\frac{1}{c^2}\big(\frac{1}{2}v^2+U\big)+O(c^{-3})$ and 
$\rho^*=\Big[1+\frac{1}{c^2}\big(\frac{1}{2}v^2+3U\big)\Big]\rho+O(c^{-4})$ at this stage. 
It is worthwhile to note that the $O(c^{-1})$ term in $T^{00}_{(1)}$ is constructed from the EMSG term that is proportional to $ d\mathfrak{M}/dt$.

We next derive the EMSG part of $T^{\mu\nu}_{\text{eff}(1)}$, i.e., the terms with the coefficient $f_0'$ in this effective tensor. We indicate this part with $T^{\mu\nu}_{\text{\tiny EMS}}$. One can show that for a perfect fluid with $L_{\text{m}}=p$, this part reduces to 
\begin{align}
\nonumber
& T^{\mu\nu}_{\text{\tiny EMS}}=f_0'\Big(c^4\rho^2g^{\mu\nu}+2c^2\big(\epsilon\rho g^{\mu\nu}+\rho^2u^{\mu}u^{\nu}\big)+\big(3p^2\\\label{T_PF_EMSG}
&+\epsilon^2\big)g^{\mu\nu}+\big(8p\rho+4\epsilon\rho+\frac{1}{c^2}(6p^2+8p\epsilon+2\epsilon^2)\big)u^{\mu}u^{\nu}\Big).
\end{align}
Utilizing the above relation and the PN expansion of the metric components, we arrive at
\begin{subequations}
\begin{align}
\nonumber
& T^{00}_{\text{\tiny EMS}(1)}={\rho^*}^2c^4f_0'\Big[1+8f_0'U_{\text{\tiny EMS}}+\frac{1}{c^2}\big(v^2+2\Pi+\frac{8p}{\rho^*}\\\label{T^00_1_EMS}
&-4U\big)\Big]+O(c^{-1}),\\\nonumber
& T^{0j}_{\text{\tiny EMS}(1)}=2{\rho^*}^2c^3f_0'
\Big[v^j\big(1+8f_0'U_{\text{\tiny EMS}}\big)-4f_0'U^j_{\text{\tiny EMS}}\\
&+\frac{2}{c^2}\Big(v^j\big(\Pi+\frac{2p}{\rho^*}-2U\big)-U^j\Big)\Big]+O(c^{-3}),\\
\nonumber
& T^{jk}_{\text{\tiny EMS}(1)}={\rho^*}^2c^4f_0'\Big[\delta^{jk}+\frac{1}{c^2}\Big(2v^jv^k-\big(v^2-2\Pi\\\label{T^jk_1_EMS}
&+8U\big)\delta^{jk}\Big)\Big]+O(c^{-2}).
\end{align}
\end{subequations}
We should mention that the $c^{-1}$ order in $T^{00}_{\text{\tiny (1)EMS}}$ is an EMSG correction and corresponds to $ d\mathfrak{M}/dt$.

To complete the source term of the gravitational potential $h_{\mathcal{N}(2)}^{\mu\nu}$, the two portions $(-g^{(1)})t_{\text{LL}(1)}^{\mu\nu}$ and $(-g^{(1)})t_{\text{H}(1)}^{\mu\nu}$ should also be calculated. Given the definitions \eqref{tLL} and \eqref{tH} and the components of $h_{\mathcal{N}(1)}^{\mu\nu}$, one can in principle obtain these portions in the near zone. We first find the PN expansion of the Landau-Lifshitz pseudotensor components below. 
\begin{subequations}
\begin{align}
\nonumber
& (-g^{(1)})t_{\text{LL}(1)}^{00}=-\frac{1}{8\pi G}\bigg[7\partial_jU\partial^jU+8c^4f_0'\Big(f_0'\partial_jU_{\text{\tiny EMS}}\\\label{t^00_1_LL}
&\times\partial^jU_{\text{\tiny EMS}}+\frac{1}{c^2}\partial_jU\partial^jU_{\text{\tiny EMS}}\Big)\bigg]+O(c^{-1}),\\\nonumber
& (-g^{(1)})t_{\text{LL}(1)}^{0j}=
\frac{1}{4\pi G c}\bigg[3\partial_tU\partial^jU+4\big(\partial^jU^k-\partial^kU^j\big)\\\nonumber
&\times\partial_kU\bigg]+\frac{c^3f_0'}{\pi G}\bigg[f_0'\big(\partial_tU_{\text{\tiny EMS}}\partial^jU_{\text{\tiny EMS}}-2\partial^jU_{\text{\tiny EMS}}\partial_kU_{\text{\tiny EMS}}^k\\\nonumber
&+2\partial^jU_{\text{\tiny EMS}}^k\partial_kU_{\text{\tiny EMS}}\big)+\frac{1}{c^2}\Big(\partial_kU_{\text{\tiny EMS}}\partial^jU^k-\partial^jU_{\text{\tiny EMS}}\partial_kU^k\\
&+2\big(\partial^jU^k_{\text{\tiny EMS}}-\partial^kU^j_{\text{\tiny EMS}}\big)\partial_kU\Big)\bigg]+O(c^{-2}),\\
\label{t^jk_1_LL}
& (-g^{(1)})t_{\text{LL}(1)}^{jk}= \frac{1}{4\pi G}\bigg[\partial^jU\partial^kU-\frac{1}{2}\partial^nU\partial_nU\delta^{jk}\bigg]\\\nonumber
&+\frac{c^4f_0'}{\pi G}\bigg[f_0'\Big(\partial^nU_{\text{\tiny EMS}}\partial_nU_{\text{\tiny EMS}}\delta^{jk}-\partial^jU_{\text{\tiny EMS}}\partial^kU_{\text{\tiny EMS}}\Big)\\\nonumber
&+\frac{1}{c^2}\Big(2\partial^jU_{\text{\tiny EMS}}\partial^kU-\partial^nU_{\text{\tiny EMS}}\partial_nU\delta^{jk}\Big)\bigg]+O(c^{-1}).
\end{align}
\end{subequations}
To obtain the PN expansion of the harmonic pseudotensor components, we substitute Eqs.  \eqref{h00_N1}-\eqref{hjk_N1} into Eq. \eqref{tH}. We have
\begin{subequations}
\begin{align}
\nonumber
&  (-g^{(1)})t_{\text{H}(1)}^{00}=-\frac{c^4f_0'}{\pi G}\bigg[f_0'U_{\text{\tiny EMS}}\partial_j\partial^jU_{\text{\tiny EMS}}+\frac{1}{c^2}U_{\text{\tiny EMS}}\partial_j\partial^jU\bigg]\\\label{t^00_1_H}
&+O(c^{-1}),\\\nonumber
&  (-g^{(1)})t_{\text{H}(1)}^{0j}=
\frac{c^3f_0'}{\pi G}\bigg[f_0'\Big(\partial_tU_{\text{\tiny EMS}}\partial^jU_{\text{\tiny EMS}}+2\partial^jU^k_{\text{\tiny EMS}}\partial_kU_{\text{\tiny EMS}}\\\nonumber
&-2U_{\text{\tiny EMS}}\partial_k\partial^kU^j_{\text{\tiny EMS}}\Big)+\frac{1}{c^2}\Big(\partial^jU_k\partial^kU_{\text{\tiny EMS}}-U_{\text{\tiny EMS}}\partial_k\partial^kU^j\Big)\bigg]\\
&+O(c^{-2}),\\
\nonumber
& (-g^{(1)})t_{\text{H}(1)}^{jk}=\frac{c^4f_0'^2}{\pi G}\bigg[\partial^jU_{\text{\tiny EMS}}\partial^kU_{\text{\tiny EMS}}-U_{\text{\tiny EMS}}\partial_n\partial^nU_{\text{\tiny EMS}}\delta^{jk}\bigg]\\\label{t^jk_1_H}
&+O(c^{-1}).
\end{align}
\end{subequations}
after some manipulations.
It should be noted that since the Landau-Lifshitz and harmonic pseudotensors are both made of the gravitational potential in the near zone, these pseudotensors are involved in the near-zone source terms in the following derivation. 

Now, we have enough information to build the components of the EMSG effective pseudotensor up to the required PN order. After gathering together Eqs. \eqref{T00_1}-\eqref{T^jk_1} and \eqref{T^00_1_EMS}-\eqref{t^jk_1_H} and some simplification, we finally arrive at
\begin{subequations}
\begin{align}
\nonumber
& c^{-2}\tau^{00}_{\text{eff}(1)}=\rho^*\bigg[1+\frac{1}{c^2}\Big(\frac{1}{2}v^2+\Pi-\frac{1}{2}U\Big)\bigg]+c^2f_0'{\rho^*}^2\\\nonumber
&\times\bigg[1+\frac{1}{c^2}\Big(v^2+2\Pi+\frac{8p}{\rho^*}-2U+\frac{2U_{\text{\tiny EMS}}}{\rho^*}\Big)\bigg]\\\nonumber
&-\frac{7}{16\pi G c^2}\nabla^2U^2-\frac{f_0'}{2\pi G}\nabla^2\big(U U_{\text{\tiny EMS}}\big)-\frac{c^2{f_0'}^2}{2\pi G}\nabla^2 U_{\text{\tiny EMS}}^2\\\label{tau_00_EMSG_1}
&+O(c^{-3}),\\\nonumber
& c^{-1}\tau^{0j}_{\text{eff}(1)}=\rho^*v^j\bigg[1+\frac{1}{c^2}\Big(\frac{1}{2}v^2+\Pi+\frac{p}{\rho^*}+3U\Big)\bigg]\\\nonumber
&+\frac{1}{4\pi G c^2}\bigg[3\partial_tU\partial^jU+4\Big(\partial^jU^k-\partial^kU^j\Big)\partial_kU\bigg]\\\nonumber
&+c^2f_0'\bigg[2{\rho^*}^2v^j+\frac{1}{c^2}\Big({\rho^*}^2v^j\big(4\Pi+\frac{p}{\rho^*}\big)-4{\rho^*}^2U^j\\\nonumber
&+\frac{1}{\pi G}\big(2\partial^jU^k\partial_k U_{\text{\tiny EMS}}-\partial^j U_{\text{\tiny EMS}}\partial_kU^k+2\big(\partial^jU^k_{\text{\tiny EMS}}\\\nonumber
&-\partial^kU^j_{\text{\tiny EMS}}\big)\partial_kU+4\pi G \rho^*v^jU_{\text{\tiny EMS}}\big)\Big)-f_0'\Big(8{\rho^*}^2U^j_{\text{\tiny EMS}}\\\label{tau_0j_EMSG_1}
&-\frac{1}{\pi G}\big(2\partial_tU_{\text{\tiny EMS}}\partial^jU_{\text{\tiny EMS}}-2\partial^jU_{\text{\tiny EMS}}\partial_kU_{\text{\tiny EMS}}^k\\\nonumber
&+4\partial^jU_{\text{\tiny EMS}}^k\partial_kU_{\text{\tiny EMS}}+8\pi G {\rho^*}^2v^jU_{\text{\tiny EMS}}\big)\Big)\bigg]+O(c^{-3}),\\
\nonumber
& \tau^{jk}_{\text{eff}(1)}=\rho^*v^jv^k+p\delta^{jk}+\frac{1}{4 \pi G}\bigg[\partial^jU\partial^kU-\frac{1}{2}\partial_nU\partial^nU\delta^{jk}\bigg]\\\nonumber
&+c^4f_0'\bigg[{\rho^*}^2\delta^{jk}+\frac{1}{c^2}\Big({\rho^*}^2\delta^{jk}\big(2\Pi-v^2-4U\big)+2{\rho^*}^2v^jv^k\\\label{tau_jk_EMSG_1}
&+\frac{1}{\pi G}\big(2\partial^jU_{\text{\tiny EMS}}\partial^kU-\partial_nU_{\text{\tiny EMS}}\partial^nU\delta^{jk}\big)\Big)\\\nonumber
&+f_0'\delta^{jk}\Big(\frac{1}{\pi G}\partial_nU_{\text{\tiny EMS}}\partial^nU_{\text{\tiny EMS}}-4{\rho^*}^2U_{\text{\tiny EMS}}\Big)\bigg]+O(c^{-1}).
\end{align}
\end{subequations}
It should be mentioned to simplify Eq. \eqref{tau_00_EMSG_1}, we use this fact that 
\begin{subequations}
\begin{align}
\label{U2}
&\nabla^2 U^2=2\partial_j U\partial^j U-8\pi G\rho^* U,\\
&\nabla^2 U_{\text{\tiny EMS}}^2=2\partial_j U_{\text{\tiny EMS}}\partial^j U_{\text{\tiny EMS}}-8\pi G{\rho^*}^2 U_{\text{\tiny EMS}},\\\nonumber
&\nabla^2\left(U U_{\text{\tiny EMS}}\right) =2\partial_j U\partial^j U_{\text{\tiny EMS}}-4\pi G{\rho^*} U_{\text{\tiny EMS}}\\
\label{UUE}
&-4\pi G{\rho^*}^2 U.
\end{align}
\end{subequations}
It is worth noting that the odd terms  $O(c^{-3})$, $O(c^{-3})$, and $O(c^{-1})$ that respectively appear in the time-time, time-space, and space-space components of $\tau^{\mu\nu}_{\text{eff}(1)}$ are entirely built of the EMSG term that is proportional to $ d\mathfrak{M}/dt$. In the following, by utilizing the near-zone source terms derived in the above relations, we attempt to find $h_{\mathcal{N}(2)}^{\alpha\beta}$.

\subsubsection*{Near-zone solution}

To find the near-zone portion of the gravitational potential, we utilize Eq. \eqref{hNear}. 
We first study the time-time component of $h_{\mathcal{N}(2)}^{\mu\nu}$. In this step, one can show that the integral \eqref{hNear} reduces to
\begin{align}
\label{h00N}
& h_{\mathcal{N}(2)}^{00}=\frac{4G}{c^2}\bigg\lbrace\int_{\mathcal{M}}\frac{c^{-2}\tau^{00}_{\text{eff}(1)}}{\rvert{\boldsymbol{x}-\boldsymbol{x}'}\rvert}d^3x'-\frac{1}{6c^3}\dddot{\mathcal{I}}^{kk}(t)\\\nonumber
&+\frac{1}{2c^2}\frac{\partial^2}{\partial t^2}\int_{\mathcal{M}}c^{-2}\tau^{00}_{\text{eff}(1)}\rvert{\boldsymbol{x}-\boldsymbol{x}'}\rvert d^3x'+\cdots\bigg\rbrace+h^{00}_{(2)}\big[\partial \mathcal{M}\big],
\end{align}
after  some simplification and applying the gauge condition which is equivalent to the conservation statement $\partial_\mu\tau^{\mu\nu}_{\text{eff}(1)}=0$ at this stage. Here, overdot stands for the derivative with respect to time $t$ and $h^{00}_{(2)}\big[\partial \mathcal{M}\big]$ shows the surface integral in this solution that actually appears after enforcing the gauge condition. In the following, we examine this term in detail. Moreover, $\mathcal{I}^{jk}$ is given by
\begin{align}
\label{Ijk}
\mathcal{I}^{jk}(t)=\int_{\mathcal{M}}c^{-2}\tau^{00}_{\text{eff}(1)}{(t,\boldsymbol{x})}x^jx^kd^3x.
\end{align}
Inserting Eq. \eqref{tau_00_EMSG_1} within Eq. \eqref{h00N}, we have
\begin{widetext}
\begin{align}
\nonumber
&h_{\mathcal{N}(2)}^{00}=\frac{4G}{c^2}\bigg\lbrace\int_{\mathcal{M}}\frac{{\rho^*}'}{\rvert{\boldsymbol{x}-\boldsymbol{x}'}\rvert}\Big(1+\frac{1}{c^2}\big(\frac{1}{2}{v'^2}+\Pi'-\frac{1}{2}U'\big)\Big)d^3x'-\frac{7}{16\pi G c^2}\int_{\mathcal{M}}\frac{\nabla'^2U'^2}{\rvert{\boldsymbol{x}-\boldsymbol{x}'}\rvert}d^3x'+c^2f_0'\int_{\mathcal{M}}\frac{{{\rho^*}'}^2}{{\rvert{\boldsymbol{x}-\boldsymbol{x}'}\rvert}}\Big(1+\frac{1}{c^2}\big(v'^2\\
\nonumber
&+2\Pi'+\frac{8p'}{{\rho^*}'}-2U'+\frac{2U'_{\text{\tiny EMS}}}{{\rho^*}'}\big)\Big)d^3x'-\frac{f_0'}{2\pi G}\int_{\mathcal{M}}\frac{\nabla'^2\big(U' U'_{\text{\tiny EMS}}\big)}{\rvert{\boldsymbol{x}-\boldsymbol{x}'}\rvert}d^3x'-\frac{c^2{f_0'}^2}{2\pi G}\int_{\mathcal{M}}\frac{\nabla'^2U'^2_{\text{\tiny EMS}}}{\rvert{\boldsymbol{x}-\boldsymbol{x}'}\rvert}d^3x'+\frac{1}{2c^2}\frac{\partial^2}{\partial t^2}\int_{\mathcal{M}}{\rho^*}'{\rvert{\boldsymbol{x}-\boldsymbol{x}'}\rvert}d^3x'\\\label{h002N}
&+\frac{f_0'}{2}\frac{\partial^2}{\partial t^2}\int_{\mathcal{M}}{{\rho^*}'}^2{\rvert{\boldsymbol{x}-\boldsymbol{x}'}\rvert}d^3x'-\frac{1}{6c^3}\dddot{\mathcal{I}}^{kk}(t)+O(c^{-4})\bigg\rbrace.
\end{align}
\end{widetext}
As seen this component is made of two portions, the compact and non-compact parts. The compact pieces are entirely constructed from the fluid variables restricted to the near zone. See the first, third, sixth, and seventh integrals in the above relation. On the other hand, the non-compact pieces can exist beyond the near zone. In fact, the sources of these terms are the Newtonian and EMSG potentials. See, the second, fourth, and fifth integrals of Eq. \eqref{h002N}. Here, we focus on these parts.

By utilizing the fact that $\nabla'^2\big(1/{\rvert{\boldsymbol{x}-\boldsymbol{x}'}\rvert}\big)=-4\pi \delta\big({\boldsymbol{x}-\boldsymbol{x}'})$, one can easily simplify the non-compact parts as follows:
\begin{subequations}
\begin{align}
\nonumber
&\int_{\mathcal{M}}\frac{\nabla'^2U'^2}{\rvert{\boldsymbol{x}-\boldsymbol{x}'}\rvert}d^3x'=-4\pi U^2+\oint_{\partial\mathcal{M}}\Big(\frac{{\partial'}^jU'^2}{\rvert{\boldsymbol{x}-\boldsymbol{x}'}\rvert}\\
&~~~~~~~~~~~~~~~~~~~~~~~-U'^2\partial'_j\frac{1}{\rvert{\boldsymbol{x}-\boldsymbol{x}'}\rvert}\Big)dS'_j,\\
&\int_{\mathcal{M}}\frac{\nabla'^2\big(U'U'_{\text{\tiny EMS}}\big)}{\rvert{\boldsymbol{x}-\boldsymbol{x}'}\rvert}d^3x'=-4\pi U U_{\text{\tiny EMS}}\\\nonumber
&~~~~+\oint_{\partial\mathcal{M}}\Big(\frac{{\partial'}^j\big(U'U'_{\text{\tiny EMS}}\big)}{\rvert{\boldsymbol{x}-\boldsymbol{x}'}\rvert}-U'U'_{\text{\tiny EMS}}\partial'_j\frac{1}{\rvert{\boldsymbol{x}-\boldsymbol{x}'}\rvert}\Big)dS'_j,\\
\nonumber
&\int_{\mathcal{M}}\frac{\nabla'^2U'^2_{\text{\tiny EMS}}}{\rvert{\boldsymbol{x}-\boldsymbol{x}'}\rvert}d^3x'=-4\pi U_{\text{\tiny EMS}}^2+\oint_{\partial\mathcal{M}}\Big(\frac{{\partial'}^jU'^2_{\text{\tiny EMS}}}{\rvert{\boldsymbol{x}-\boldsymbol{x}'}\rvert}\\
&~~~~~~~~~~~~~~~~~~~~~~~-U'^2_{\text{\tiny EMS}}\partial'_j\frac{1}{\rvert{\boldsymbol{x}-\boldsymbol{x}'}\rvert}\Big)dS'_j,
\end{align} 
\end{subequations}
in which $d S_j=\mathcal{R}_j^2\sin\theta\, d\theta\, d\varphi$. 
Considering the definitions of the standard and EMSG potentials, one can also deduce that $U\propto 1/\mathcal{R}$ and $U_{\text{\tiny EMS}}\propto 1/\mathcal{R}$ on the boundary $\partial \mathcal{M}$. See Appendix D of \cite{nazari2020Constraining}. So all surface integrals in the above relations will be proportional to $\propto 1/\mathcal{R}^2$. In the framework of the modern approach to the PN approximation, it is claimed that the $\mathcal{R}$-dependent can be removed during calculations. In fact, it is argued that the $\mathcal{R}$-dependent terms in the near-zone and wave-zone solutions will eventually be canceled by each other \cite{poisson2014gravity}. We use this scheme here and drop all surface integrals. Given these points, we then simplify the time-time component as
\begin{align}
\nonumber
& h_{\mathcal{N}(2)}^{00}=\frac{4U}{c^2}+4f'_0U_{\text{\tiny EMS}}+\frac{1}{c^4}\bigg\lbrace 4\psi+7U^2+2\partial_{tt}X\\\nonumber
&+4c^2f'_0\Big(\psi_{\text{\tiny EMS}}+2 UU_{\text{\tiny EMS}}+2c^2f'_0U^2_{\text{\tiny EMS}}+\frac{1}{2}\partial_{tt} X_{\text{\tiny EMS}}\Big)\bigg\rbrace\\\label{h002}
&-\frac{2G}{3c^5}\dddot{\mathcal{I}}^{kk}(t)+O(c^{-5}),
\end{align}
in which
\begin{subequations}
\begin{align}
&\psi=G\int_{\mathcal{M}}\frac{{\rho^*}'}{{\rvert{\boldsymbol{x}-\boldsymbol{x}'}\rvert}}\Big(\frac{1}{2}v'^2+\Pi'-\frac{1}{2}U'\Big)d^3x',\\
& X=G\int_{\mathcal{M}}{{\rho^*}'}\rvert{\boldsymbol{x}-\boldsymbol{x}'}\rvert d^3x',
\end{align}
\end{subequations}
are the well-known PN potentials and
\begin{align}
\nonumber
&\psi_{\text{\tiny EMS}}= G\int_{\mathcal{M}}\frac{{{\rho^*}'}^2}{{\rvert{\boldsymbol{x}-\boldsymbol{x}'}\rvert}}\Big(v'^2+2\Pi'+\frac{8p'}{{\rho^*}'}-2U'\\
&~~~~~~~~~~~~~~~~~~~~~~~~~~~~~~~+\frac{2U'_{\text{\tiny EMS}}}{{\rho^*}'}\Big)d^3x',\\
\label{XEMS}
& X_{\text{\tiny EMS}}=G\int_{\mathcal{M}}{{\rho^*}'}^2\rvert{\boldsymbol{x}-\boldsymbol{x}'}\rvert d^3x',
\end{align}
are the new gravitational potentials defined in the PN limit of the EMSG theory. For the sake of simplification, let us break down $\psi_{\text{\tiny EMS}}$ in terms of several EMSG potentials as follows: 
\begin{align}
\psi_{\text{\tiny EMS}}=V_{\text{\tiny EMS}}+2\Pi_{\text{\tiny EMS}}+8P_{\text{\tiny EMS}}-2\,\mathcal{U}^{(1)}_{\text{\tiny EMS}}+2\,\mathcal{U}^{(2)}_{\text{\tiny EMS}},
\end{align}
where 
\begin{subequations}
\begin{align}
& V_{\text{\tiny EMS}}=G\int_{\mathcal{M}}\frac{{{\rho^*}'}^2v'^2}{\rvert{\boldsymbol{x}-\boldsymbol{x}'}\rvert}d^3x',\\\label{Pi}
& \Pi_{\text{\tiny EMS}}=G\int_{\mathcal{M}}\frac{{{\rho^*}'}^2\Pi'}{\rvert{\boldsymbol{x}-\boldsymbol{x}'}\rvert}d^3x',\\
& P_{\text{\tiny EMS}}=G\int_{\mathcal{M}}\frac{{\rho^*}'p'}{\rvert{\boldsymbol{x}-\boldsymbol{x}'}\rvert}d^3x',
\end{align}
\end{subequations}
as well as
\begin{subequations}
\begin{align}
\label{U1}
&\mathcal{U}^{(1)}_{\text{\tiny EMS}}=G\int_{\mathcal{M}}\frac{{{\rho^*}'}^2U'}{\rvert{\boldsymbol{x}-\boldsymbol{x}'}\rvert}d^3x',\\
\label{U22}
&\mathcal{U}^{(2)}_{\text{\tiny EMS}}=G\int_{\mathcal{M}}\frac{{{\rho^*}'}U'_{\text{\tiny EMS}}}{\rvert{\boldsymbol{x}-\boldsymbol{x}'}\rvert}d^3x'.
\end{align}
\end{subequations}

Regarding Eqs. \eqref{g00}-\eqref{gjk}, the other component of the gravitational potential required to build the metric is $h^{kk}$.
To find this component, we need to obtain its source term, i.e., $\tau_{\text{eff}}^{kk}$. Considering Eq. \eqref{tau_jk_EMSG_1}, we have
\begin{align}
\nonumber
&\tau_{\text{eff}(1)}^{kk}=\rho^*\big(v^2-\frac{1}{2}U\big)+3p-\frac{1}{16\pi G}\nabla^2U^2\\
\nonumber
&+c^2f'_0\bigg[3{\rho^*}^2c^2\Big(1+\frac{1}{c^2}\big(2\Pi-\frac{1}{3}v^2-\frac{14}{3}U-\frac{2}{3}\frac{U_{\text{\tiny EMS}}}{\rho^*}\big)\Big)\\\label{tau_kk_EMSG_1}
&-\frac{1}{2\pi G}\nabla^2\big(UU_{\text{\tiny EMS}}\big)+\frac{3c^2f'_0}{2\pi G}\nabla^2U^2_{\text{\tiny EMS}}\bigg]+O(c^{-1}),
\end{align}
in which Eqs. \eqref{U2}-\eqref{UUE} are inserted. Substitution of this source term into the integral 
\begin{align}
\label{hkkN}
& h_{\mathcal{N}(2)}^{kk}=\frac{4G}{c^4}\bigg\lbrace\int_{\mathcal{M}}\frac{\tau^{kk}_{(1)\text{eff}}}{\rvert{\boldsymbol{x}-\boldsymbol{x}'}\rvert}d^3x'-\frac{1}{2c}\dddot{\mathcal{I}}^{kk}(t)\\\nonumber
&+\frac{1}{2c^2}\frac{\partial^2}{\partial t^2}\int_{\mathcal{M}}\tau^{kk}_{(1)\text{eff}}\rvert{\boldsymbol{x}-\boldsymbol{x}'}\rvert d^3x'+\cdots\bigg\rbrace+h^{kk}_{(2)}\big[\partial \mathcal{M}\big],
\end{align}
gives rise to
\begin{align}
\nonumber
& h_{\mathcal{N}(2)}^{kk}=12 f'_0U_{\text{\tiny EMS}}+\frac{1}{c^4}\bigg\lbrace 4V+U^2+24c^2f'_0\Big(\Pi_{\text{\tiny EMS}}\\
\nonumber
&-\frac{1}{6}V_{\text{\tiny EMS}}-\frac{7}{3}\,\mathcal{U}^{(1)}_{\text{\tiny EMS}}-\mathcal{U}^{(2)}_{\text{\tiny EMS}}+\frac{1}{3}UU_{\text{\tiny EMS}}+\frac{1}{4}\partial_{tt}X_{\text{\tiny EMS}}\\\label{hkk2}
&-c^2f'_0U^2_{\text{\tiny EMS}}\Big)\bigg\rbrace-\frac{2G}{c^5}\dddot{\mathcal{I}}^{kk}(t)+O(c^{-5}).
\end{align}
Here, 
\begin{align}
V=G\int_{\mathcal{M}}\frac{{\rho^*}'}{\rvert{\boldsymbol{x}-\boldsymbol{x}'}\rvert}\Big(v'^2-\frac{1}{2}U'+\frac{3p'}{{\rho^*}'}\Big)d^3x',
\end{align}
is another PN potential. It is worth mentioning that Eq. \eqref{hkkN} is obtained after inserting the conservation statement $\partial_\mu\tau^{\mu\nu}_{\text{eff}(1)}=0$ within the general near-zone solution \eqref{hNear}.

Finally, for the time-space and space-space components of the gravitational potential, in a similar fashion to the previous calculation, we arrive at
\begin{align}
\label{h0j2}
 h_{\mathcal{N}(2)}^{0j}=\frac{4}{c^3}\bigg\lbrace U^j+2c^2f'_0U^j_{\text{\tiny EMS}}\bigg\rbrace+O(c^{-5}),
\end{align}
and
\begin{align}
\nonumber
& h_{\mathcal{N}(2)}^{jk}=4f'_0\delta^{jk}U_{\text{\tiny EMS}}+\frac{4}{c^4}\bigg\lbrace  W^{jk}+X^{jk}+\frac{1}{4}\delta^{jk}U^2\\\nonumber
&+c^2f'_0\Big[\delta^{jk}\big(2\Pi_{\text{\tiny EMS}}-V_{\text{\tiny EMS}}-6\,\mathcal{U}^{(1)}_{\text{\tiny EMS}}-2\,\mathcal{U}^{(2)}_{\text{\tiny EMS}}+2UU_{\text{\tiny EMS}}\\\nonumber
&+\frac{1}{2}\partial_{tt}X_{\text{\tiny EMS}}-2c^2f'_0U^2_{\text{\tiny EMS}}\big)+2U^{jk}_{\text{\tiny EMS}}+8X^{jk}_{\text{\tiny EMS}}\Big]\bigg\rbrace\\\label{eq3}
&-\frac{2G}{c^5}\dddot{\mathcal{I}}^{jk}(t)+O(c^{-5}),
\end{align}
in which the standard tensorial potentials $W^{jk}$ and $X^{jk}$ are given by
\begin{subequations}
\begin{align}
& W^{jk}=G\int_{\mathcal{M}}\frac{{\rho^*}'}{\rvert{\boldsymbol{x}-\boldsymbol{x}'}\rvert}\Big(v'^jv'^k-\frac{U'}{2}\delta^{jk}+\frac{p'}{{\rho^*}'}\delta^{jk}\Big)d^3x',\\
& X^{jk}=\frac{1}{4\pi}\int_{\mathcal{M}}\frac{\partial^jU'\partial^kU'}{\rvert{\boldsymbol{x}-\boldsymbol{x}'}\rvert}d^3x',
\end{align}
\end{subequations}
respectively. The EMSG tensorial potentials $U^{jk}_{\text{\tiny EMS}}$ and $X^{jk}_{\text{\tiny EMS}}$ are also defined as
\begin{subequations}
\begin{align}
& U^{jk}_{\text{\tiny EMS}}=G\int_{\mathcal{M}}\frac{{{\rho^*}'}^2v'^jv'^k}{\rvert{\boldsymbol{x}-\boldsymbol{x}'}\rvert}d^3x',\\
& X^{jk}_{\text{\tiny EMS}}=\frac{1}{4\pi}\int_{\mathcal{M}}\frac{\partial^jU'_{\text{\tiny EMS}}\partial^kU'}{\rvert{\boldsymbol{x}-\boldsymbol{x}'}\rvert}d^3x'.
\end{align}
\end{subequations}

As the final point at this stage, let us focus on the surface terms that are involved in the integrals \eqref{h00N} and \eqref{hkkN} and of course in the general integral of $h_{\mathcal{N}(2)}^{0j}$ and $h_{\mathcal{N}(2)}^{jk}$. We return to Eq. \eqref{hNear} before enforcing the conservation statement $\partial_\mu\tau^{\mu\nu}_{\text{eff}(1)}=0$. The second term in this expansion has an important role because it may turn to the surface integrals in the leading PN orders. Setting $l=1$, we have
\begin{align}
\label{eq2}
-\frac{1}{c}\frac{d}{dt}\int_{\mathcal{M}}\tau^{\mu\nu}_{\text{eff}(1)}(t,\boldsymbol{x})d^3x,
\end{align}
for the second term in the PN expansion of $h_{\mathcal{N}(2)}^{\mu\nu}$.
Now, we evaluate the role of this term in the PN expansion of each component of the gravitational potential.
For the time-time component, using the conservation equation $\partial_0\tau^{00}_{\text{eff}(1)}+\partial_j\tau^{j0}_{\text{eff}(1)}=0$, one can easily show that this integral is simplified as $\oint_{\partial\mathcal{M}}\tau^{0j}_{\text{eff}(1)}dS_j$. Considering the slow-motion condition, we find that the compact parts of $\tau^{0j}_{\text{eff}(1)}$ do not exist at the boundary of the region $\mathcal{M}$ and consequently their surface integrals vanish. Moreover, regarding Eq. \eqref{tau_0j_EMSG_1}, the surface integral of the non-compact pieces is proportional to $1/\mathcal{R}$ and $1/\mathcal{R}^2$. So, this part of the integral is $\mathcal{R}$-dependent, and it can be discarded freely. 
To simplify the time-space component of Eq. \eqref{eq2}, we also use the space component of the conservation equation, i.e., $\partial_0\tau^{0j}_{\text{eff}(1)}+\partial_k\tau^{jk}_{\text{eff}(1)}=0$. In this case, this integral is reduced to $\oint_{\partial\mathcal{M}}\tau^{jk}_{\text{eff}(1)}dS_k$. In a similar way, we conclude that the compact pieces of Eq. \eqref{tau_jk_EMSG_1} have no role in this surface integral and the non-compact ones all are proportional to $1/\mathcal{R}^2$.
To examine the role of Eq. \eqref{eq2} in the last component of the potential, i.e., $h_{\mathcal{N}(2)}^{jk}$, we need to do more calculations. Here, this integral is given by $-\partial_0\int_{\mathcal{M}}\tau^{jk}_{\text{eff}(1)}(t,\boldsymbol{x})d^3x$. To simplify this integral, we use the identity \cite{poisson2014gravity}
\begin{align}
\tau^{jk}_{\text{eff}}=\frac{1}{2}\partial_{00}\big(\tau^{00}_{\text{eff}}x^jx^k\big)+\frac{1}{2}\partial_p\big(\tau^{pj}_{\text{eff}}x^k+\tau^{pk}_{\text{eff}}x^j-\partial_q\tau^{pq}_{\text{eff}}x^jx^k\big),
\end{align}
deduced from the conservation equations. By inserting this relation into the integral, after some simplification, we arrive at $-1/(2c) \dddot{\mathcal{I}}^{jk}-1/2\oint_{\partial\mathcal{M}}\big(\tau^{pj}_{\text{eff}(1)}x^k+\tau^{pk}_{\text{eff}(1)}x^j-\partial_q\tau^{pq}_{\text{eff}(1)}x^jx^k\big)dS_p$. Utilizing Eq. \eqref{tau_jk_EMSG_1}, we find that the non-compact pieces of the surface integral all are proportional to $1/\mathcal{R}$ and as before, the compact parts are zero. So, the contribution of Eq. \eqref{eq2} to the PN expansion of $h_{\mathcal{N}(2)}^{jk}$ lies in $-1/(2c) \dddot{\mathcal{I}}^{jk}$. We exhibit this term in Eq. \eqref{eq3}.
In conclusion, the non-zero surface integrals coming from this leading term in PN expansion of the potential, all are $\mathcal{R}$-dependent and they can be dropped.
It can be shown that the other surface integrals from the next PN terms in this expansion play a role in the higher PN corrections and do not appear in the 1\tiny PN \normalsize order.
Therefore, up to the required PN order in this work, the surface integrals have no contribution to the components of $h_{\mathcal{N}(2)}^{\alpha\beta}$ and we remove them from Eqs. \eqref{h002}, \eqref{hkk2}, \eqref{h0j2}, and \eqref{eq3}.

\subsubsection*{wave-zone portion}

Up to this point, we have obtained the near-zone portion of the gravitational potential in the second iteration. To complete our derivation, we need to find its wave-zone part, $h^{jk}_{\mathcal{W}(2)}$. 
The source terms of this potential are comprehensively introduced in \cite{nazari2020Constraining}. 
Let us rewrite these terms here.
\begin{subequations}
\begin{align}
\nonumber
&\tau_{\text{eff}(1)}^{00}=-\frac{G}{\pi r^4}\Big(\frac{7}{8}M_0^2+3 c^4f_0'\mathfrak{M}\big(f_0'\mathfrak{M}+\frac{1}{c^2}M_0\big)\Big)\\\label{tauW00}
&~~~~~~~~~~+O(c^{-1}),\\\label{tauW0j}
& \tau_{\text{eff}(1)}^{0j}=O(c^{-1}),\\
\nonumber
& \tau_{\text{eff}(1)}^{jk}=\frac{G}{\pi r^4}\Big[\Big(\frac{1}{4}M_0^2+2c^2f_0'\mathfrak{M}M_0\Big)\Big(n^jn^k-\frac{1}{2}\delta^{jk}\Big)\\\label{eq6}
&~~~~~~~~~~-c^4{f_0'}^2\mathfrak{M}^2\delta^{jk}\Big]+O(c^{-1}).
\end{align}
\end{subequations}
In the above relations, $M_0$ given by
\begin{align}
M_0=\int_{\mathcal{M}}{\rho^*}d^3x',
\end{align}
is the total matter inside the near-zone region. These PN expansions of $\tau_{\text{eff}(1)}^{\mu\nu}$ are sufficient to construct the wave-zone solution of the gravitational potentials in the second iteration.

According to Eq. \eqref{eq8}, to evaluate this integral, one should first rewrite the source functions in the form of Eq. \eqref{eq7}.  
Comparing Eq. \eqref{tauW00} with Eq. \eqref{eq7} reveals that $n=4$, $l=0$, and 
\begin{align}
f^{00}_{l=0}=-4G\Big[\frac{7}{8}M_0^2+3c^2f'_0\mathfrak{M}\big(M_0+c^2f'_0\mathfrak{M}\big)\Big].
\end{align}
Therefore, for the time-time component, the $f^{\alpha\beta}(\tau)$ function is constant. Knowing that $n=4$, one can then easily derive the integral \eqref{eq8} as
\begin{align}
h_{\mathcal{W}(2)}^{00}=\frac{2G f_{l=0}^{00}}{c^4}\frac{1}{\mathcal{R}^2}.
\end{align}
As seen, this component of the potential is a function of $\mathcal{R}$.
For $h_{\mathcal{W}(2)}^{0j}$, given Eqs. \eqref{tauW0j} and \eqref{eq8}, one can show that the source term of this potential does not construct the required $O(c^{-3})$ correction for this component. So, $h_{\mathcal{W}(2)}^{0j}$ does not contribute to the 1\tiny PN \normalsize order of the metric.
Similarly, to build $h_{\mathcal{W}(2)}^{jk}$, we simplify the source term of this potential, i.e., Eq. \eqref{eq6}, as follows:
\begin{align}
\nonumber
&\tau_{\text{eff}(1)}^{jk}=\frac{G}{\pi r^4}\bigg[\Big(\frac{1}{4}M_0^2+2c^2f_0'\mathfrak{M}M_0\Big)\Big(n^{<jk>}\\\label{eq11}
&~~~~~~~~~-\frac{1}{6}\delta^{jk}\Big)-c^4{f_0'}^2\mathfrak{M}^2\delta^{jk}\bigg]+O(c^{-1}),
\end{align}
where $n^{<jk>}=n^jn^k-1/3\,\delta^{jk}$.
Regarding Eqs. \eqref{eq7} and \eqref{eq11}, one can then show that
\begin{subequations}
\begin{align}
\label{eq9}
& f_{l=0}^{jk}=-\frac{G}{6}\delta^{jk}\Big(M_0^2+8c^2f_0'\mathfrak{M}\big(3 c^2f_0'\mathfrak{M}+ M_0\big)\Big),\\
\label{eq10}
& f_{l=2}^{jk}=G M_0\big(M_0+8c^2f_0'\mathfrak{M}\big),
\end{align}
\end{subequations}
and $n=4$ in this case.
Regarding the these points, we have
\begin{subequations}
\begin{align}
& h_{\mathcal{W}(2)}^{jk}\vert_{l=0}=\frac{2G f^{jk}_{l=0}}{c^4}\frac{1}{\mathcal{R}^2},\\
& h_{\mathcal{W}(2)}^{jk}\vert_{l=2}=\frac{G f^{jk}_{l=2}}{5 c^4}n^{<jk>}\frac{r^2}{\mathcal{R}^4}.
\end{align}
\end{subequations}
It is seen that the wave-zone terms calculated here all are a function of $\mathcal{R}$. As we have learned from the modern approach, we can drop these  $\mathcal{R}$-dependent terms freely. Therefore, in the second iteration, the wave-zone portion of the gravitational potential has no role in our calculations, and  we finally have $h_{(2)}^{\mu\nu}=h_{\mathcal{N}(2)}^{\mu\nu}$.

Our final item of business in this subsection is to survey the importance of the odd PN orders $c^{-5}$ that may appear in the time-time component of the metric. To do so, let us build this component. Inserting Eqs. \eqref{h002} and \eqref{hkk2} within Eq. \eqref{g00}, we arrive at
\begin{align}
\nonumber
& g^{(2)}_{00}=-1+\frac{2}{c^2}U+\frac{2}{c^4}\Big(\psi+V-U^2+\frac{1}{2}\partial_{tt} X\Big)\\\nonumber
&+4f'_0\Big[2U_{\text{\tiny EMS}}+\frac{1}{c^2}\Big(\partial_{tt}X_{\text{\tiny EMS}}+4P_{\text{\tiny EMS}}-4UU_{\text{\tiny EMS}}\\\nonumber
&-8\,\mathcal{U}^{(1)}_{\text{\tiny EMS}}-2\,\mathcal{U}^{(2)}_{\text{\tiny EMS}}+4\Pi_{\text{\tiny EMS}}\Big)-11f'_0U^2_{\text{\tiny EMS}}\Big]\\
&-\frac{4G}{3c^5}\dddot{\mathcal{I}}^{kk}(t)+O(c^{-5}).
\end{align}
The first concerning term is $\dddot{\mathcal{I}}^{kk}(t)$ which is only a function of time. It is shown this type of terms can be removed by applying an appropriate time coordinate transformation \cite{poisson2014gravity}. To examine the impact of the other odd orders collected in $O(c^{-5})$, let us return once more to the general expansion of the metric in terms of the gravitational potentials. As grasped from Eq. \eqref{g00} for $g_{00}$, 
$h^{00}$ and $h^{kk}$ are the two main factors that can bring this odd order into this expression. 
By considering several leading PN terms of Eq. \eqref{hNear} that can have a role in this order and also regarding the origin of the $c^{-1}$ terms in the PN expansions of $\tau^{00}_{\text{eff}(1)}$ and $\tau^{kk}_{\text{eff}(1)}$,
we deduce that most of the $O(c^{-5})$ terms from these potentials are made entirely by the EMSG term which corresponds to $(c^{-3}f'_0+c^{-1}{f'_0}^2) d\mathfrak{M}/dt$ multiplied by terms like $\partial_k U$ and $\partial_kU_{\text{\tiny EMS}}$.
We should emphasize that although $d\mathfrak{M}/dt$ itself is only a function of time, the other coefficients, such as $\partial_k U$ and $\partial_kU_{\text{\tiny EMS}}$, are a function of $x$ and $t$. Then, we cannot treat this $O(c^{-5})$ term as the previous correction and remove it by a transformation of the time coordinate.
On the other hand, we show in Appx. \ref{Newtonian hydrodynamics} that $c^2f'_0d\mathfrak{M}/dt$ is of the order $c^{-2}$.  Hence, these apparently $O(c^{-5})$ terms actually contribute to the order $c^{-7}$.
The next odd-order term we should treat carefully, comes from the unusual high-order term, $3c^4f'_0{\rho^*}^2$, in $\tau_{\text{eff}(1)}^{kk}$. 
This term is of the order $c^{2}$. So, it can produce a multipole moment of the order $c^{-5}$ for $l=3$ in the expansion \eqref{hNear} for $h_{\mathcal{N}(2)}^{kk}$. Dropping the numerical coefficients and constant parameters, this so-called multipole moment contributes to $h_{\mathcal{N}(2)}^{kk}$ as 
$\partial_{ttt}\int_{\mathcal{M}}{\rho^*}^2(t,\boldsymbol{x}) x^2d^3x$. As seen, after integrating, this term indeed depends only on time. Therefore, it is a coordinate artifact and it can be omitted along with $\dddot{\mathcal{I}}^{kk}(t)$ by applying a suitable time coordinate transformation. However, in general, the $O(c^{-7})$ terms may produce during this transformation \cite{poisson2014gravity}. So, using this transformation just removes the odd power $c^{-5}$ and the next PN corrections with the odd power $c^{-7}$ may appear in the PN expansion of the metric.
Eventually, we deduce that the terms with the order $c^{-5}$ appearing in the PN expansion of $g_{00}^{(2)}$ all have a role in the next PN corrections and they are actually of the order $c^{-7}$.
It is worth mentioning that the $O(c^{-7})$ corrections, GR terms and especially those coming from EMSG terms, can involve the odd numbers of time derivatives. Consequently, like GR, this approximate solution of the field equation is not invariant under the time reflection up to this order. This fact illustrates that the system described by this metric loses energy via radiating gravitational waves. As shown, several EMSG terms exist within this order. So, they can in principle contribute to gravitational waves emitted from this system. The role of the EMSG corrections in the gravitational waves is studied in \cite{nazari2020Constraining}.

\section{Order of magnitude of $d\mathfrak{M}/dt$}\label{Newtonian hydrodynamics}

In this appendix, we survey the order of magnitude of $d \mathfrak{M}/dt$ which appears frequently in our calculation. 
To do so, we utilize the local conservation equations derived during the second iteration, i.e., $\partial_{\mu}\tau^{\mu\nu}_{\text{eff}(1)}=0$.
Considering the time component of $\partial_{\mu}\tau^{\mu\nu}_{\text{eff}(1)}=0$, we have
\begin{align}
\partial_0\tau^{00}_{\text{eff}(1)}+\partial_k\tau^{0k}_{\text{eff}(1)}=0,
\end{align}
and then inserting Eqs. \eqref{tau_00_EMSG_1} and \eqref{tau_0j_EMSG_1} within this relation, we find that
\begin{align}
\label{eq12}
\partial_{t}\rho^*+\partial_{j}\big(\rho^*v^j\big)+2c^2f'_0\Big(\rho^*\partial_t\rho^*+\partial_j\big({\rho^*}^2v^j\big)\Big)=O(c^{-2}).
\end{align}
We have also assumed that the conservation of rest-mass, $\nabla_{\mu}\big(\rho u^{\mu}\big)=0$, is established in EMSG. $\rho^*$ then satisfies the continuity equation
\begin{align}\label{rho}
\partial_t\rho^*+\partial_j(\rho^*v^j)=0.
\end{align}
Applying Eq. \eqref{rho} for the Newtonian and EMSG sectors of Eq. \eqref{eq12}, we deduce that
\begin{align}
\label{rule}
2c^2f'_0 \rho^* v^j\partial_j{\rho^*}=O(c^{-2}).
\end{align}
So, the energy conservation statement reveals that this term is of the order $c^{-2}$. 
On the other hand, in appendix C of \cite{nazari2020Constraining}, we have shown that $d \mathfrak{M}/dt$ is proportional to the integration
of $\rho^* v^j\partial_j \rho^*$. Given Eq. \eqref{rule}, one can deduce that $c^2f'_0d \mathfrak{M}/dt$ is indeed of the order $c^{-2}$. We use this fact to indicate the PN order of some odd-power terms in the PN expansion of the metric in Appx. \ref{app_1}.

\end{document}